# Cleaner Production in Optimized Multivariate Networks: Operations Management through a "Roll of Dice"

**Amit K Chattopadhyay**[a,*]
[a]Aston University, Systems Analytics Research Group, Mathematics, Birmingham B4 7ET, UK

*Corresponding Author: Email: a.k.chattopadhyay@aston.ac.uk

**Biswajit Debnath**[b,c]
[b]Department of Chemical Engineering, Jadavpur University, Kolkata 700032, India
[c]Aston University, System Analytics Research Group, Birmingham B4 7ET, UK

Email: biswajit.debnath.ju@gmail.com

**Rihab El-Hassani**[d]
[d]ENSIIE – Ecole Nationale Suprieure d'Informatique pour l'Industrie et l'Enterprise, Paris, Evry, France
Email: rihab.elhassani@gmail.com

**Sadhan Kumar Ghosh**[e]
[e]Department of Mechanical Engineering, Jadavpur University, Kolkata 700032, India
Email: sadhankghosh9@gmail.com

**Rahul Baidya**[e]
[e]Department of Mechanical Engineering, Jadavpur University, Kolkata 700032, India
Email: rahulbaidya.ju@gmail.com

**Abstract**
The importance of supply chain management in analyzing and later catalyzing economic expectations while simultaneously prioritizing cleaner production aspects is a vital component of modern finance. Such predictions, though, are often known to be less than accurate due to the ubiquitous uncertainty plaguing most business decisions. Starting from a multi-dimensional cost function defining the sustainability of the supply chain (SC) kernel, this article outlines a 4-component SC module - environmental, demand, economic, and social uncertainties – each ranked according to its individual weight. Our mathematical model then assesses the viability of a sustainable business by first ranking the potentially stochastic variables in order of their subjective importance, and then optimizing the cost kernel, defined from a 'utility function'. The model will then identify conditions (as equations) validating the sustainability of a business venture. The ranking is initially obtained from an Analytical Hierarchical Process; the resultant "weighted cost function" is then optimized to analyze the impact of market uncertainty based on our supply chain model. Model predictions are then ratified against SME data to emphasize the importance of cleaner production in business strategies.

## 1. Introduction

In today's competitive business world, sustainability is a synonym for rapid adaptation to changes occurring economically, politically and socially. There is a paradigm shift towards 'sustainability-practice' approach rather than 'sustainability-performance' approach (Silva and Figueiredo 2020). While the economics of profitability is a key business driver, sustainability needs to be ingrained for its long-termed survival incorporating both ethical and environmental sustenance. Changes in global business performance are increasingly being monitored using Sustainable Development Goals (SDG) indicators, like how green the supply chain is, carbon rating, energy efficiency, gender equity, etc. (unstats.un.org). Fluctuations in these indicators affect the equilibrium of the supply chain network, often leading to fluctuating business performance. These fluctuating performances effectively amount to time evolving perturbations in their respective time series data. Such perturbations in turn amount to performance uncertainty that affect the consumer supply chain dynamics with potentials for key impact on the operations management of supply chain (Fathollahi-Fard et al. 2018).

The success of modern-day business is highly dependent on the efficiency of a supply chain in adapting to these uncertainties such that even in situations with large fluctuations, internal micro-management is capable of identifying "performance windows" within which the supply line can still be sustained. In other words, it is imperative that for a business willing to operate in a 'sustainability-practice' approach should be pre-advised about such working windows beyond which the supply chain logistics should not be allowed to vacillate, both for the purpose of sustainability as also for economic development. As of 2015, the Paris Agreement adopted by 193 countries in the 21st Session of Conference of the Parties (COP21) stipulates global carbon emission to be reduced as much as possible to keep the global temperature increase below a 1.5$^{o}$C (UNFCCC 2015). The United Nations' Inter-Governmental Panel on Climate Change (IPCC) adjudication stipulates carbon credit as a requirement and not an option, an aspect that has a knock-on effect on the uncertainty of business performances, especially for SMEs (UNFCCC 1998). The backdrop demands 'cleaner' supply chains with simultaneous higher levels of 'production'.

### *1.1 Supply Chain Sustainability and the concept of Cleaner Production (CP)*

Supply Chain Network or simply Supply Chain on its own is a phrase that represents a complex structure with numerous major and minor loop leading towards product manufacturing, distribution and disposal. Ideally, a supply chain network (SCN) consists of three sections – Supply Side, Internal Operations and Demand Side (Mandal 2015). Although Supply Side and Demand Side both incur internal fluctuations, hence uncertainties, Internal operations are more critical in cleaner sustenance of the Supply Chain adhering sustainability requirements as they confine the boundaries of the production system within the value chain. The production systems hence need to be designed in such a way that they offer high productivity and economic efficiency simultaneously.

In line with the discussion above, identifying waste producing processes and/or inefficient supply chain kernels in a setup is of paramount importance as that can affect the production-sustainability dyad directly (Mandal 2015). This requires economically sustainable and environmentally cleaner technologies inspiring green supply chain and innovative strategizing (Van Berkel 2002). The concept of Cleaner Production (CP) is strategically based on the internal operations of a Supply Chain Network (SCN). According to UNEP (1997), CP can be defined as "*the continuous application of an integrated preventative environmental strategy to processes and products to reduce risks to humans and the environment.*" CP is largely understood as the bridging strategy between sustainable development and waste prevention (Van Berkel 2002; Van Berkel 2007). It not only includes resource efficient utilization of raw materials but also conservation of energy, elimination of hazardous substances and reduction of quantity and toxicity of wastes (Van Berkel 2007). However, CP is not strictly environmental, rather it provides an optimization platform between economic growth (Economic), environmental protection (Environmental), resource efficiency (Demand) and social equity (Van Berkel 2002).

Traditional semantics is based on three basic pillars of sustainability, yet these can be logically customized to four or more pillars conforming to the evolving definition of sustainability, as the boundaries of the concept are not always explicitly defined (Schlör et al. 2015). Reported literature shows that the fourth pillar could be Institutional (Dawodu et al. 2017), operational (Debnath & Ghosh 2019) and/or cultural (Soini and Birkeland 2014) based on system and situation. Since our focus is an SCN with CP, our model uses a 4-dimensional sustainability array, respectively environmental, economic, social and demand. Demand plays a crucial role in the supply chain and even the slightest market volatility affects the demand which in turn affects the sustainability of the supply chain. As mentioned at the start, modern SCN networks rely more on ‘sustainability-practice’ making it imperative to identify as well as quantify any inherent upcoming risk. As an example, the four identified SCN pillars have several minute operational nodes which can pose serious threat to the supply chain as well as to the business logistics. As discussed by Van Berkel (2002), the concept of CP is intertwined with these four pillars offering enough scope to incorporate the appropriate elements of CP. The agenda then is to develop best prevention strategies that can reduce uncertainty braided risk in devising CP strategies.

The Sustainable Development Goals (SDGs) 2030 were adopted by the member countries of United Nations in the year 2015. The target was to strategically end poverty, protect natural resources and ensure political tranquility by 2030. The SDGs can be achieved by controlling the supply chain uncertainties towards its own benefits by implementing CP strategies (Figure 1). The following subsections outline supply chain uncertainties and their connection with SDGs.

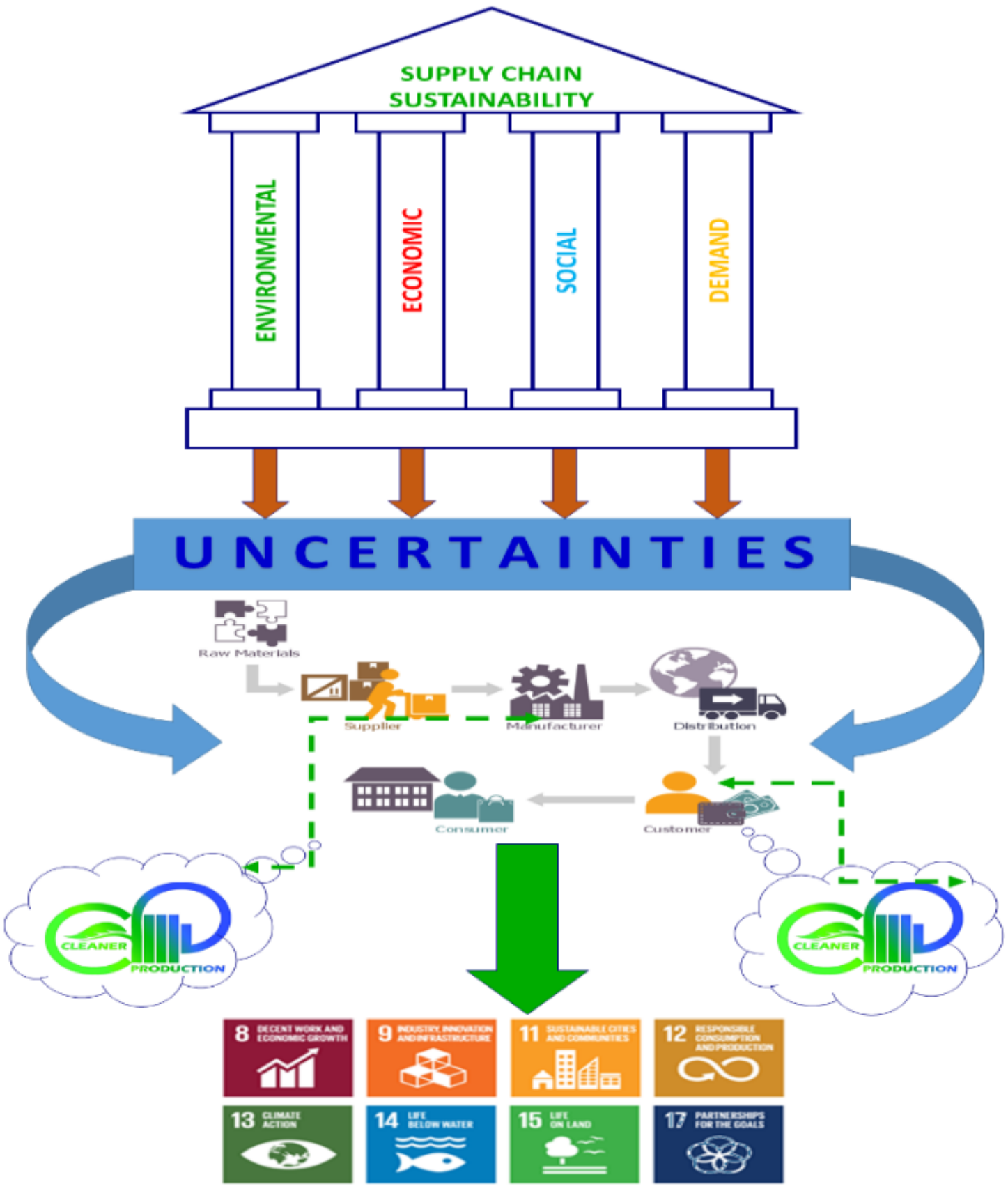

Figure 1: Interrelationship of Supply Chain Uncertainties, Cleaner Production and SDGs (The pillars show the supply chain sustainability dimensions, the brown arrows signify the contribution of the four pillars towards uncertainties, the blue arrows signify uncertainties affecting SCN, cloud callouts linked with double-sided arrows suggest that connection of CP strategies is a feedback process similar to a PDCA cycle. The Green Arrow links the cleaner and greener output leading towards addressing SDGs 8,9,11-15 and 17.

### *1.2 Supply Chain Uncertainty and Sustainable Development Goals (SDGs)*

Uncertainties in a SCN can be routed back to rapid dynamic changes occurring at different circumstances and environmental conditions of a production-redistribution-disbursement triad (Hund et al. 2001). These are likely to have unwanted and/or unexpected perturbations due to economical and sociological situations, affecting the supply chain network. The nature of these perturbations is often unfamiliar and may or may not be time variant.

In general, the reported literature shows uncertainty in different aspects – their role, effects, risks, network framework, product design, system design, firm performance, models, supply chain flexibility, sensitivity analysis, investment planning, green supply chain practices, strategic decision making and optimization. It was mentioned in the preceding subsection that each of the four pillars of supply chain sustainability has certain aspects which can induce perturbations in the form of uncertainties. Existing literature suggest that there are four type of uncertainties – a) Environmental uncertainties (Patel et al, 2012); b) Demand uncertainties (Kim et al. 2018); c) Social uncertainties (Fathollahi-Fard et al. 2018) and d) Economic uncertainties (Zhang et al. 2011).

In contemporary literature, uncertainty has generally been considered as a parameter in supply chain modeling in the last decade (Patel et al. 2012; Cardoso et al. 2013). For most cases, uncertainties can exist in different forms and patterns. From the users' end, this can be taxonomically re-oriented. A large body of works dwell primarily on demand uncertainty (Cardoso et al. 2013; Kim et al. 2018). Cardoso et al. 2013 developed a Mixed Integer Linear Programming (MILP) model for designing and planning of supply chain and reverse logistics considering demand uncertainty using a scenario tree approach. Kim et al. (2018) developed a deterministic mixed-integer optimization model with additional counterparts that are robust enough to cope with the uncertainty of recycled products and customer demand in the fashion industry. Hence, considering demand as the fourth pillar of supply chain sustainability is meaningful and counterfeits the claim with the compendium of literature available on demand uncertainty. In relations to the SDGs, goal no. 7, 9, 11 and 12 can be addressed through manipulation of demand uncertainty.

Supply chain literature is replete with examples of environmental uncertainties in supply chain modeling (Patel et al, 2012). The concept of environmental uncertainty dates back to Thompson (Thompson 1967) and the pioneering studies on environmental uncertainty (Snyder, 1987**),** or perceptually (Lorenzi et al., 1981), or as a combination of both (Milliken 1987). Previous literature acknowledges the effects of environmental uncertainty in the supply chain network (Tachizawa and Thomsen, 2007; Vickery et al., 1999). However, the results outlined in these publications are not all uniform, exhibiting both negative impact (Han et al. 2014), positive impact (Patel et al. 2012) and some unchanged (Pagell and Krause 2004). It is important to note that these studies focus on corporate environment rather than on environmental factors e.g. carbon emission, waste product treatment, etc. affecting the supply chain network specifically. This provides a scope of intervention to actually consider the environmental factors, e.g. SDGs 6, 7, 12, 13, 14 and 15 (Figure 1) as environmental achievable.

Only limited literature is available on economic and social uncertainty. Fathollahi-Fard et al. (2018) developed a multi-objective stochastic programming model that combines the effects of both economic and social uncertainties in a closed loop supply chain network. While literature is suffused with deterministic MILP modules, fuzzy mathematics was also employed to optimize uncertainty against system variables (Pishvaee and Razmi 2012). Zhang et al. (2011) demonstrates a multi-echelon, multi-product supply chain production planning model based on multiple stochastic kernels was found effective for solving the problem. SDGs 1-5, 8-11, 16 and 17 are the socio-economic goals that comprise several aspects including zero hunger, gender equality, good health, industry-innovation-infrastructure and sustainable cities, to name a few. Not all of these goals can be addressed but a topical few, closely related to economics and development, are important. Given the fact that SDGs are largely generic in their remit, combining supply chain uncertainties with CP strategies is a challenging task, that we address in this work.

### *1.3 Existing Knowledge Gaps*

As defined at the beginning of the preceding subsection, uncertainties are responsible both for the origin and outcome of market fluctuations. Mathematically, they are represented as stochastic fluctuations, both positive (profit line) and negative (loss line), in relevant SDG market (model) predictions, effects of which are replicated at different nodes of the supply chain network. It is quite an important and challenging task for the supply chain managers to deal with the uncertainties and maintain a sustainable business in practice.

A number of solutions have been proposed by researchers to tackle SDG uncertainty issues in supply chain. Typical examples include (but are not limited to) fuzzy mathematics (Pishvaee and Razmi 2012); robust optimization (Kim et al. 2018); MILP (Cardoso et al. 2013); stochastic programming (Xie and Huang 2018) etc. It is also interesting that demand uncertainty has been a long-time favorite for the researchers (Kim et al. 2018) while attention on other uncertainties are slowly emerging. On the other hand, supply chain sustainability has already been taken up by a lot of researchers (Mota et al. 2015; Chowdhury et al. 2020), but most of them fail to combine multiple parameters within a holistic description (Chowdhury et al. 2020). These wide ranging multitudinous forms of uncertainty (stochasticity) clearly suggest a *knowledge gap in integrating the impact of the time evolution of uncertainty with the eventual model prediction*.

The present work targets to bridge this incoherence by constructing a time dynamical model of costing in which market or subjective uncertainty interacts directly with cost minimization kernels, work-routine optimization and eventual profit maximization in an interconnected nebular supply-chain network. We incorporate uncertainties across the entire breadth of the supply chain network that are likely to have profound impact on the business sustainability, especially when two or more variables are affected simultaneously. As discussed, existing literature fails to capture the effects of (all four) uncertainties collectively in a single model, together with a distinctive lack of weighing their relative importance in a specific supply-chain kernel. Our study first unifies all four types of uncertainties, with a focus on the 4 pillars of sustainability, in a single cost function-based model and then focuses on the effects of individual variables affecting the supply chain network in an unconstrained and a constrained environment, leading to a prioritization subroutine of respective uncertainties in order of their importance in that supply chain for cleaner production.

## 2. Problem Statement and Model Framework

### *2.1 Problem Statement*

While it is well known that the role of uncertainty is a key decision maker in the profitability aspect and even the survival probability of a company (Li and Hu 2014), the effects are even more so with Small and Medium Enterprises (SMEs) that operate on shoe string budgets on which a minor fluctuation could cause a major flutter. In other words, a model of an SME could serve as a litmus test of the veracity of any uncertainty model proposed. This is a major motivation for this study. Additionally, we consider four pillars of sustainability, including the three basic pillars (environmental, economic and social) and an additional fourth pillar quantifying demand uncertainty.

SMEs often struggle to adapt to the perturbations induced by the uncertainty variables and it is of vital importance to collectively consider the uncertainties in decision making. As mentioned before, the supply chain uncertainties affect the sustainability and tune them within controllable limits to drive cleaner production. Hence, it is also important to prioritize the (four) phenomenological sources of uncertainty and include the effects of the correlation of these variables into a single model. The goal of this paper is to develop a mathematical model incorporating these four uncertainties and the effects of their interdependencies with the remits of a single model. The emergent solution toolbox will rank the uncertainty parameters/variables, quantify their respective and collective contributions in the market dialectic, and identify parameter windows that ensure their sustenance and profitability regimes.

### *2.2 Model Description*

#### *2.2.1 Assumptions*

The assumptions considered in this model are stated as follows –

i) All calculations are on daily basis that is yearly data to be divided by 3000, assuming 10 hours' work for each working day, over 300 working days in a year.
ii) Cost associated with the legislation and miscellaneous cost remains constant.
iii) Unit costs remain constant.
iv) Profit (per unit) is taken to be constant.

#### *2.2.2 The Free Energy Model*

The four uncertainties has been considered to develop the model which is structured around a cost function F defined as a linear summation of the four uncertainty functions (Eq.1). This function is the first step towards quantification of subjective uncertainty. A "cost function" is itself, can be interpreted as a "free energy" (Baker 2000). Every uncertainty function is a linear combination of three or more identified variables contributing towards uncertainty (Eq. 2 – 5). The environmental uncertainty function is a function of different environmental parameters that calls for uncertainty e.g. Water consumed, wastewater generated etc. (Eq.2). The social uncertainty function is a linear combination of the uncertainty of the social parameters considered (Eq.3).

Economic uncertainty can be measured as the difference of profit from the sum of depleting costs, like Operating cost, cost of disaster management and taxes, each of which is stochastic in its remit (Eq.4). The demand uncertainty is defined as linear summation of Cost of Transportation, Cost of other Logistics & Packaging and other miscellaneous costs (Eq.5). The named variables span all four dimensions of supply chain sustainability and they are critically chosen such that it is possible to monitor each major and minor perturbations along the supply chain. Essentially, these variables are not only connected to pollution aspects but also to several other socio-economic features and demand facets which, chosen by the model, will open the regime for implementation of cleaner production strategies. The sources of the variables are available in the Supplementary document (Table S3).

$$F = C_{Environment} + C_{Social} + C_{Economic} + C_{Demand} \tag{1}$$

where

$$C_{Environment} = \sum_i V_{CO_2} f_{1_i} + \sum_i H_{P_i} f_{2_i} + \sum_i W_{P_i} f_{3_i} + \sum_i W_{w_i} y_i + \sum_i L_i \tag{2}$$

$$C_{Social} = \sum_i N_{1_i} f_{4_i} + \sum_i N_{2_i} f_{5_i} + \sum_i N_{3_i} f_{6_i} \tag{3}$$

$$C_{Economic} = \sum_i N_{4_i} f_{7_i} - \sum_i N_{5_i} f_{8_i} - \sum_i f_{9_i} g_i - \sum_i T_i N_{6_i} \tag{4}$$

$$C_{Demand} = \sum_i f_{10_i} N_{7_i} + \sum_i f_{11_i} N_{8_i} + \sum_i M_i \tag{5}$$

Together with the relevant (AHP) weight factors, as detailed later, the free energy structured then reads as follows:

$$F = \varepsilon_1 \left( \sum_i A_1 V_{CO_2} f_{1_i} + \sum_i A_2 H_{P_i} f_{2_i} + \sum_i A_3 W_{P_i} f_{3_i} + \sum_i A_4 W_{w_i} y_i + \sum A_5 L_i \right) + \varepsilon_2 \left( \sum A_6 N_{1_i} f_{4_i} + \sum_i A_7 N_{2_i} f_{5_i} + \sum_i A_8 N_{3_i} f_{6_i} \right)$$
$$+ \varepsilon_3 \left( \sum_i A_9 N_{4_i} f_{7_i} - \sum_i A_{10} N_{5_i} f_{8_i} - \sum_i A_{11} f_{9_i} g_i - \sum_i A_{12} T_i N_{6_i} \right) + \varepsilon_4 \left( \sum A_{13} f_{10_i} N_{7_i} + \sum_i A_{14} f_{11_i} N_{8_i} + \sum_i A_{15} M_i \right) \tag{6}$$

Here $\epsilon_1, \epsilon_2, \epsilon_3, \epsilon_4$ respectively define the weights corresponding to the individual uncertainties. In absence of real market data, these too have been derived through AHP analysis.

*2.2.3 Inter-dependency of the parameters*

The variables considered in this paper are often inter-dependent, requiring parametric multivariate calculus with interaction terms. To project outcomes at the linear level, each such mutually dependent variable has been expressed as a combination of the dependent variables that are either directly or indirectly inspired by the need to address cleaner production while simultaneously optimizing profit lines. We assume quadratic order accuracy (Nelder 1977).

$$V_{CO_2} = V_{CO_2}(L, N_5, N_7) = a_1 L + a_2 N_5 + a_3 N_7 + a_{12} L N_5 + a_{23} N_5 N_7 + a_{31} L N_7 + a_1' L^2 + a_2' {N_5}^2 + a_3' {N_7}^2 \tag{7a}$$

$$W_P = W_P(N_5) = W_P^O + b_1 N_5 + b_2 {N_5}^2 \tag{7b}$$

$$H_P = H_P(L, N_5) = c_1 L + c_2 N_5 + c_{12} L N_5 + c_1' L^2 + c_2' {N_5}^2 \tag{7c}$$

$$W_w = W_w(L, N_5) = d_1 L + d_2 N_5 + d_{12} L N_5 + d_1' L^2 + d_2' {N_5}^2 \tag{7d}$$

$$N_3 = N_3\left(N_4, N_6\right) = \alpha_1 N_4 + \alpha_2 N_6 + \alpha_{12} N_4 N_6 + \alpha_1^{'} {N_4}^2 + \alpha_2^{'} {N_5}^2 \tag{7e}$$

$$N_4 = N_4\left(N_7, N_8\right) = \beta_1 N_7 + \beta_2 N_8 + \beta_{12} N_7 N_8 + \beta_1^{'} {N_7}^2 + \beta_2^{'} {N_8}^2 \tag{7f}$$

$$N_7 = N_7(V_{CO_2}) = \gamma V_{CO_2} \tag{7g}$$

**3. Methods & Methodology**

**3.1 Modeling**

The model is inspired by the four uncertainty categories pertaining to the SCN of any company which are identified via literature review, brainstorming and case studies (Ergan et al. 2010; Kim et al. 2018). The developed 'free energy' model is a linear cost function-based model and the core structure has been discussed in detail in section 2.2.2. Weight factors are introduced into the 'free energy' model which represents proportional magnitudes of contribution from each variable towards uncertainty. This increases the robustness of the model as the effects of each variable will be distributed in a realistic manner. These variables are not independent of each other and their interrelations have profound effect on the supply chain dynamics. Section 2.2.3 provides further details on it.

In order to figure out the weight factors and values of the co-efficient of the interrelationships of the variables parameters, Analytical Hierarchical Process (AHP) is used (Chowdhury et al. 2018; Vishwakarma et al. 2019). Two AHPs are used for the aforementioned purposes – the first one is a simple AHP which is used to prioritize the uncertainty variables and the second one is a layered AHP (with two layers of alternatives) which has been used to prioritize the inter-dependence parameters. The eigenvalues obtained from the AHP have been used to derive the weight factors. The ratings of the AHP have been obtained from an Indian SME (anonymous).

The 'free energy' model is then converted into an optimization problem to optimize the cost function. The resultant hessian matrix represents the general scenario of a supply chain network highlighting the possible sources of uncertainty (Wang et al 2019). The optimization runs through two gateways – (a) Unconstrained Optimization (Zhu et al. 2018) and (b) Constrained Optimization (Quddus et al. 2018). The unconstrained optimization problem deals with the largely hypothetical scenario in which the supply chain has unlimited resource, both in finance, work force and supply lines, whereas the constrained optimization problem portrays the realistic scenario of a company subjectively limited by its resource and targets. The unconstrained optimization is dealt using a Hamiltonian Process (Goldstein 1964; Hamill 2014) whereas the constrained optimization process is solved by introducing constraints through Lagrange Multipliers (Elton et al. 2009). The models are validated using data obtained from an anonymous Indian SME.

The following flowchart outlines the problem to solution stream:

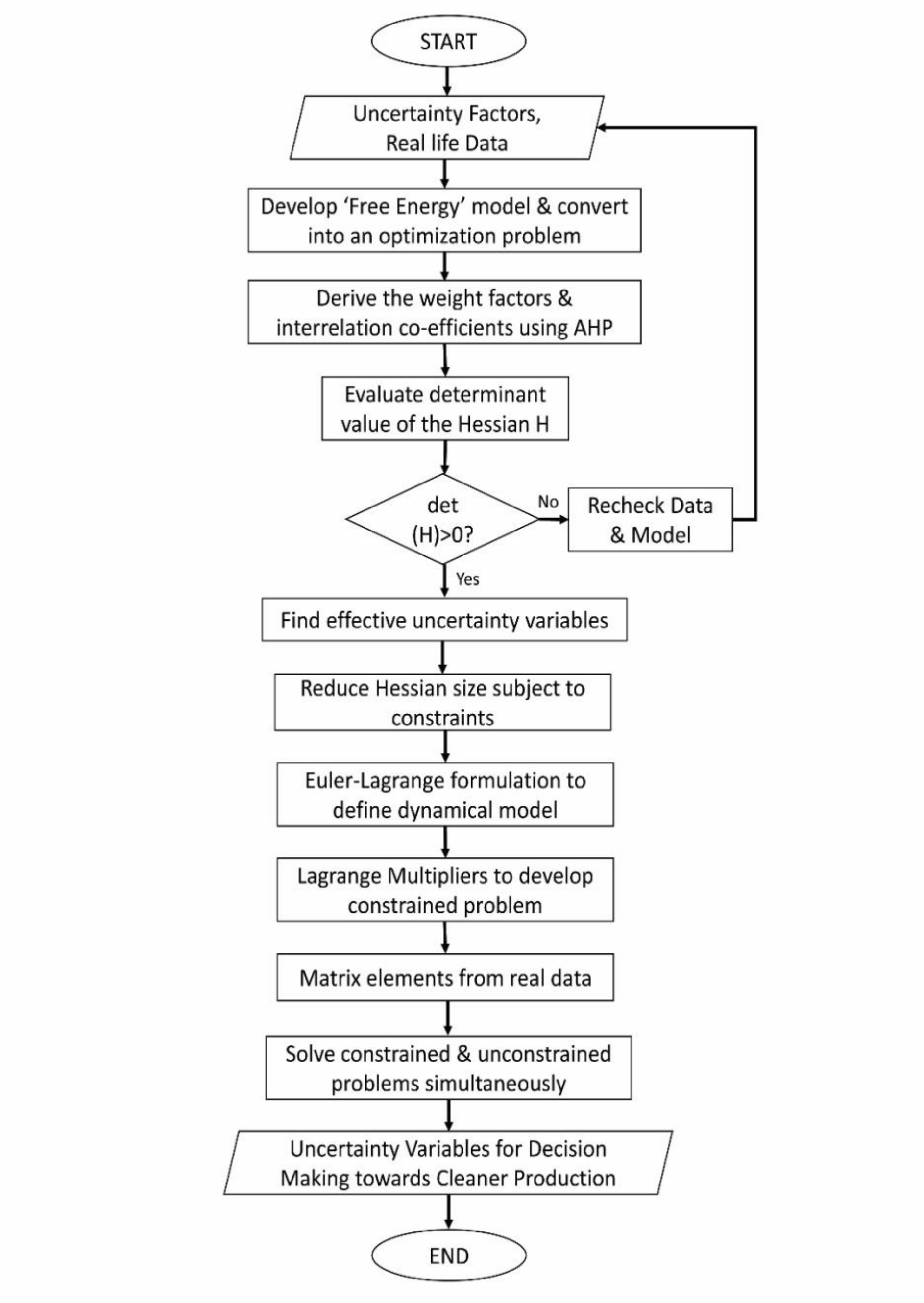

Figure 2: Working Flowchart of the solution approach

### 3.2. Solution

### 3.2.1 Analytical Hierarchical Process

Decision making involves multiple criteria and sub-criteria that are used to rank all possible outcomes of a decision. Not only does a supply chain need to prioritize its end-deliverables and future target lines, but also it should provision itself for inherent alternatives (Vishwakarma et al. 2019). In line with the existing literature, AHP has been used to route the deterministic components of our supply chain variables, primarily using the "Super Decision software" (http://sdbeta.superdecisions.com/), the operational algorithm of which, as coined by Saaty (Saaty 1980), is elaborated through a running flowchart in the appendix 1.

### 3.2.2 AHP models

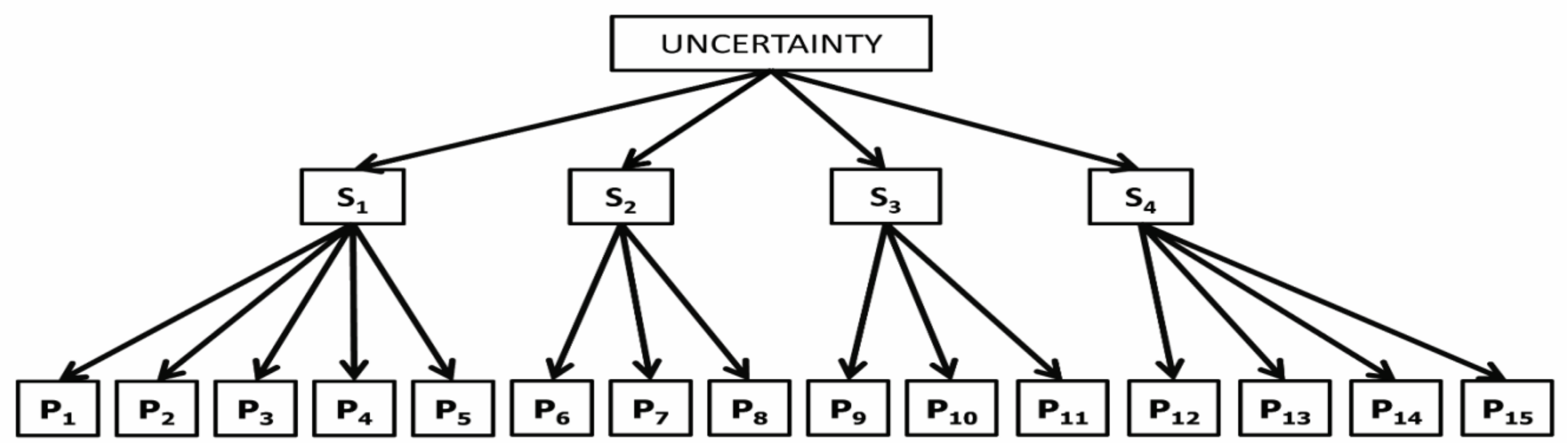

Figure 3: AHP model 1 to determine the general alterative rankings

In this paper, two complementary AHP analyzes were carried out, the first of which enumerated the weights of the variables concerned and the follow-up AHP estimated the weights of the inter-dependent variables (linearized) of each of these initial set of variables. The first is a hierarchical structure that gives the alternative ($P_1$ to $P_{15}$) rankings (i.e. the concerned parameters for the uncertainties); this has been used as weights/coefficients ($A_1$ to $A_{15}$) of the objective function F. It consists of a goal cluster, a criteria cluster and an alternatives cluster (Figure 3). The second AHP is a layered diamond like hierarchical structure. It consists of a goal cluster, a criteria cluster and two layered alternatives (Figure 4). The layered structure enables to connect the interdependent parameters in a simple and easier way. Detailed description is presented in Supplementary material (Appendix 1).

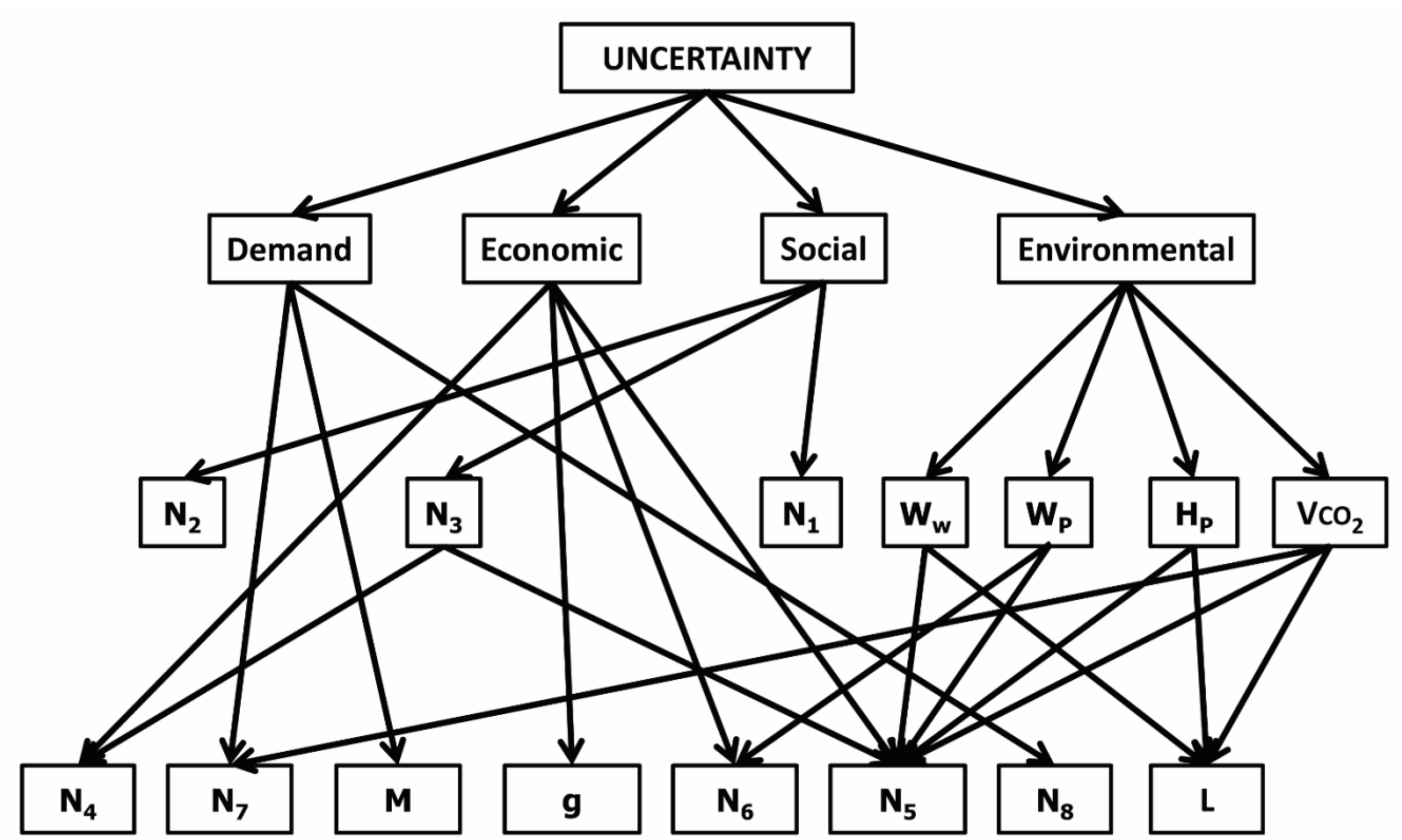

Figure 4: Layered AHP model for determination of interrelationship values

### 3.2.3 Unconstrained Problem

In reality, not all of the uncertainty parameters will induce equal levels of uncertainty in the system. As an example, we take the constrained case of an Indian SME characterized against the number of products sold ($N_4$); number of operations involved ($N_5$); number of mode of transportations ($N_7$) and volume of carbon dioxide emission (VCO2). These parameters were perturbed in order to check the fluctuation of the deterministic time depending free energy aka the cost function.

As is expected in popular macroeconomic dynamics, the time dependent movements of the cost function modules are expected to abide the multivariate Euler-Lagrange structure (Goldstein 1964) $\delta\left(\frac{\partial F}{\partial N_i}\right)=\delta\left(\frac{d}{dt}\frac{\partial F}{\partial \acute{N_i}}\right)$ that, in terms of the leading dynamical variables $(N_4, N_5, N_7, VCO_2)$, then leads to the following dynamical system (details in the Appendix)

$$\delta\left(\frac{d}{dt}\begin{bmatrix}\frac{\partial F}{\partial N_4}\\ \frac{\partial F}{\partial N_5}\\ \frac{\partial F}{\partial N_3}\\ \frac{\partial F}{\partial N_1}\\ \frac{\partial F}{\partial N_2}\\ \frac{\partial F}{\partial N_6}\\ \frac{\partial F}{\partial N_7}\\ \frac{\partial F}{\partial N_8}\\ \frac{\partial F}{\partial V_{CO_2}}\\ \frac{\partial F}{\partial L}\end{bmatrix}\right)=\begin{bmatrix}x_{1,1} & 0 & 0 & 0 & 0 & 0 & x_{1,7} & x_{1,8} & 0 & 0\\ 0 & x_{2,2} & 0 & 0 & 0 & 0 & x_{2,7} & 0 & 0 & x_{2,10}\\ x_{3,1} & 0 & x_{3,3} & 0 & 0 & x_{3,6} & 0 & 0 & 0 & 0\\ 0 & 0 & 0 & 0 & 0 & 0 & 0 & 0 & 0 & 0\\ 0 & 0 & 0 & 0 & 0 & 0 & 0 & 0 & 0 & 0\\ x_{6,1} & 0 & 0 & 0 & 0 & x_{6,6} & 0 & 0 & 0 & 0\\ 0 & x_{7,2} & 0 & 0 & 0 & 0 & x_{7,7} & x_{7,8} & 0 & x_{7,10}\\ 0 & 0 & 0 & 0 & 0 & 0 & x_{8,7} & x_{8,8} & 0 & 0\\ 0 & x_{9,2} & 0 & 0 & 0 & x_{9,6} & 0 & 0 & x_{9,9} & x_{9,10}\\ 0 & x_{10,2} & 0 & 0 & 0 & 0 & x_{10,7} & 0 & 0 & x_{10,10}\end{bmatrix}\begin{bmatrix}\delta N_4\\ \delta N_5\\ \delta N_2\\ \delta N_1\\ \delta N_2\\ \delta N_6\\ \delta N_7\\ \delta N_8\\ \delta V_{CO_2}\\ \delta L\end{bmatrix} \qquad (8)$$

This leads to

$$\delta\left(\frac{d}{dt}\begin{bmatrix}\frac{\partial F}{\partial N_4}\\ \frac{\partial F}{\partial N_5}\\ \frac{\partial F}{\partial N_7}\\ \frac{\partial F}{\partial V_{CO_2}}\end{bmatrix}\right)=\begin{bmatrix}m_{11} & 0 & m_{13} & 0\\ 0 & m_{22} & m_{23} & 0\\ 0 & m_{32} & m_{33} & 0\\ 0 & m_{42} & 0 & m_{44}\end{bmatrix}\begin{bmatrix}\delta N_4\\ \delta N_5\\ \delta N_7\\ \delta V_{CO_2}\end{bmatrix} \qquad (9)$$

$$\frac{d^2}{dt^2}\begin{bmatrix}\xi_1\delta N_4\\ \xi_2\delta N_5\\ \xi_3\delta N_7\\ \xi_4\delta V_{CO_2}\end{bmatrix}=\begin{bmatrix}m_{11} & 0 & m_{13} & 0\\ 0 & m_{22} & m_{23} & 0\\ 0 & m_{32} & m_{33} & 0\\ 0 & m_{42} & 0 & m_{44}\end{bmatrix}\begin{bmatrix}\delta N_4\\ \delta N_5\\ \delta N_7\\ \delta V_{CO_2}\end{bmatrix}$$

(10)

As may have already been spotted by a mathematically intensive reader, Eq. (9) represents a Euler-Lagrange formulation (Goldstein 1964) leading to an 'under-damped' model in Eq. (10) (Risken 1996).

### 3.2.4 Constrained Problem

A supply chain is not defined through a generic cost function; rather it is subjectively defined only with respect to the constraints that it is practically subjected to, such as – economic limit; restrictions on number & quality of work forces; transport restrictions etc. To analyze such constrained problems, a Lagrangian structure for the cost function has been used where constraints were introduced through Lagrange multipliers (Goldstein 1964) as it allows the optimization problem to be solved without explicit parameterization in terms of the constraints (Tur et al. 2009). The structure allocates weight factors to individual Lagrange multipliers which are assumed to be proportional to the epsilon values replicating the respective weightage of the uncertainty in the cost factors. The Lagrangian 'L' is defined as –

$$L = F - \lambda_1(N_1 f_4 + N_2 f_5 - C) - \lambda_2(N_4 f_7 - E) - \lambda_3\left(V_{CO_2} f_1 - V\right) - \lambda_4(N_3 f_6 - R), \quad (11)$$

where the $\lambda_i$'s are the Lagrange multipliers. The actual system restrictions, identified by the quantities coupled with the coefficient $\lambda$'s, define the (four) constraints that we enforce on the system. These pertain to our choice of the SME data to be analyzed shortly. In real terms, the quantity C represents the maximum budget accorded for the wages of the laborers and employees; E represents the total earning expected based on products sold; V is the cost related to $CO_2$ control; R identifies the maximum allowable CSR cost. Overall, this provides an equitable quantification of a 'cleaner' production line.

The following is the constrained form of the problem

$$\delta\left(\frac{d}{dt}\begin{bmatrix}\frac{\partial L}{\partial N_4}\\ \frac{\partial L}{\partial N_5}\\ \frac{\partial L}{\partial N_7}\\ \frac{\partial L}{\partial V_{CO_2}}\end{bmatrix}\right)=\begin{bmatrix}k_{11} & 0 & k_{13} & 0\\ 0 & k_{22} & k_{23} & 0\\ 0 & k_{32} & k_{33} & 0\\ 0 & k_{42} & 0 & k_{44}\end{bmatrix}\begin{bmatrix}\delta N_4\\ \delta N_5\\ \delta N_7\\ \delta V_{CO_2}\end{bmatrix}$$

(12)

$$\frac{d^2}{dt^2}\begin{bmatrix}\psi_1\delta N_4\\ \psi_2\delta N_5\\ \psi_3\delta N_7\\ \psi_4\delta V_{CO_2}\end{bmatrix}=\begin{bmatrix}k_{11} & 0 & k_{13} & 0\\ 0 & k_{22} & k_{23} & 0\\ 0 & k_{32} & k_{33} & 0\\ 0 & k_{42} & 0 & k_{44}\end{bmatrix}\begin{bmatrix}\delta N_4\\ \delta N_5\\ \delta N_7\\ \delta V_{CO_2}\end{bmatrix}$$

(13)

Equations (10) and (13) are solved using data obtained from an anonymous Indian SME company. Independent verification is attained using MATLAB R2018b (bvp4c) and Mathematica 12 (structured Runge-Kutta 4) concerning the solutions of the corresponding boundary value problems.

## 4. Results and Discussions

In this section, the behaviors of most important parameters with respect to time in constrained (utopian) as well as unconstrained system (dystopian) environment have been elucidated. All constrained parameters are identified through extra 'c's alongside the main variable; e.g. $Nc_4$ i.e. product sold in constrained environment. In reality, several occasions may arise where the SME has to devise specific strategies to control their business. Such strategies can be mathematically represented by appropriate Initial Conditions (IC) and Boundary Conditions (BC) defining the time varying system.

This study considers two different strategies that a SME can take represented in the form of boundary conditions (Table 2). The successive subsections present two case studies which reflect the effect of decision making considering the uncertainty variables contribute in flourishing business and/or bankruptcy. Ours will be a minimalist approach where maximally varying variables/parameters will be identified and then analyzed. As to less fluctuating variables (i.e. more stable variables), at this level, we tacitly subsume their fluctuation dependence of the cost function. The main variables sensitive to this system are the number of products sold ($N_4$) and volume of carbon dioxide emission ($VCO_2$).

As in the case previously discussed, we analyze the carbon footprint of an SME as a function of varying cost functions. This is done to analyze what could be the best optimized strategy of an SME to cope with such environmental demands against their stringent budgets. In terms of our model, the effects of the correlation (specific to a certain case and may vary with other companies) of $N_4$ and $VCO_2$ on business growth will be really interesting as this will reflect the best $CO_2$ minimization-versus-profit optimization scenario. The trade-off between these two rival variables will define the optimization strategy by identifying an optimal time point when the SME can flourish while also ensuring environmental sustainability.

Table 2: Boundary Conditions for Evaluation of Results

| **Sl.** | **$N_4$** | **$N_5$** | **$N_7$** | **$VCO_2$** | **$VCO_2$ Check** | |
|---|---|---|---|---|---|---|
| | | | | | **3-Y** | **5-Y** |
| | | | ***IC*** | | | |
| 1 | 0.3 | 76 | 2 | 2.86 | 3 | 2.3 |
| 2 | 0.3 | 76 | 2 | 2.86 | 2.19 | 1.46 |
| | | | ***BC*** | | | |
| 1 | 0.6 | 80 | 5 | 5.2 | 2.4 | 2 |
| 2 | 0.5 | 82 | $\frac{dN_7}{dt} = 0$ | 2 | 1.72 | 1 |

### 4.1 *Case Study 1: The SME chooses to increase the product sale without compromising on the carbon reduction (i.e. carbon emission increases)*

This is a classic case of ignorance where the SME chooses to increase the production and aims to flourish by increasing sale. However, they are economically constrained so as not to be able to invest any extra fund to address the issue of growing $CO_2$ emission. As a result, the boundary conditions change not only for these two variables but also the transportation cost increases. This case complements SDG 9, 13-15 i.e. industry, innovation and infrastructure; climate action, life on land and life under water.

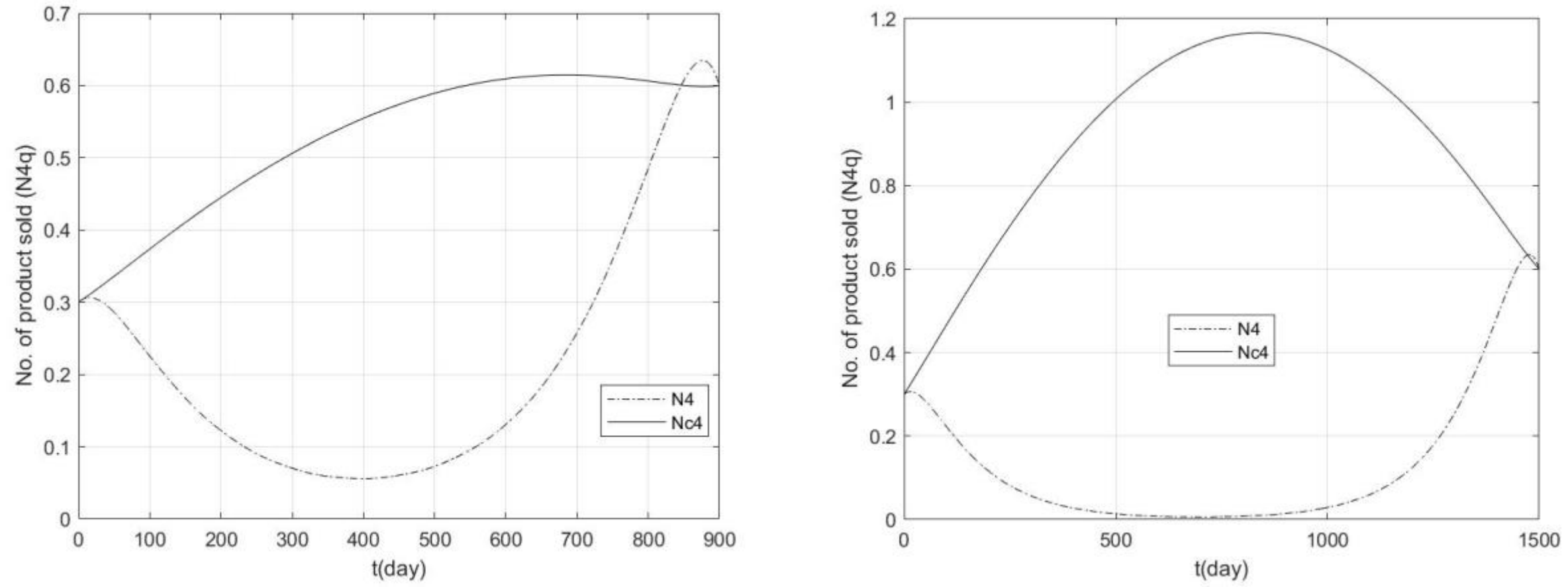

(a) (b)

Figure 5: Time dependency of $N_4$ in unconstrained (dotted line) and constrained (solid line) environment, plots obtained from simultaneous solution of Eqs. (10) and (13): (a) 3-year time span; (b) 5-year time span

In the unconstrained environment, the company starts to lose money and reaches its minimum in the first quarter of the second year (Figures 5a). Eventually, it gains momentum by raising their sale and exceeds the target slightly and then reaches the goal by end of the third year. However, in the 5-year time span (Figures 5b), the curve flattens and then rises after the third year indicating that even in the most favorable of conditions, the strategy will not give the SME best results in the long run. The realistic version of the results shows that in a 3-year time frame, the company smoothly matches its projected target within the stipulated timeline. In the longer run (5-year time frame), our model predicts that this company will start depreciating and will only break even after the second year. However, it rises to its pinnacle towards the end of the third year as it exceeds the target. Thereafter, the curve steeply falls to reach its target value.

In a theoretical world allowing for investment over an infinitely large time period (10+ years) of investment, this strategy will fail to give good results as the curve will become oscillatory and hence unstable. The results clearly indicate that the rise and fall in sales are due to demand uncertainty and the SME needs to take further actions to flourish, a predictive perspective that we arrive at from our model analysis. An inbuilt assumption to all such analyzes is the zero-uncertainty enforcement at the end points of a boundary value system, a situation that mathematically mimics a certainty in a decision before a process starts and after it ends, a logical paradigm. As suggested by the outcome, the change is strategy is the key which is well connected to SDG 9.

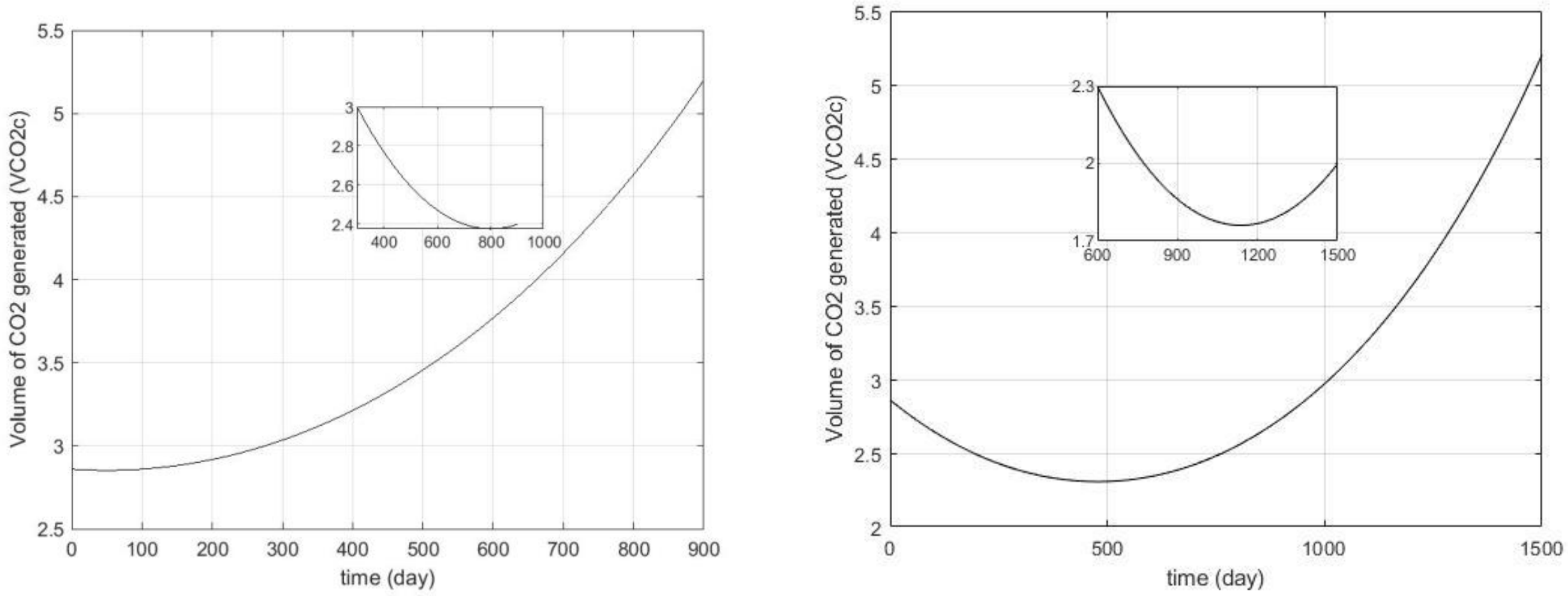

(a) (b)

Figure 6: Time Dependency of $V_{CO2}$ in a constrained environment, plots obtained from simultaneous solution of Eqs. (10) and (13): (a) 3-year time span & (b) 5-year time span

In the unconstrained environment, $VCO_2$ gives non-converging solutions which imply that without carbon emission control, the system is unstable. The sample curve for the unconstrained environment is provided in the Appendix 3. As expected, the emission curve rises to its peak value in a parabolic profile at both 3 (Figures 6a) and 5-year (Figures 6b) timelines, implying that the adopted strategy is not environmentally sustainable. *To address this, our model advises the company to re-strategize the carbon footprint at the end of each relevant time period, identified by the minimum in the respective plots*. The results are shown in the insets of figure 6(a) and 6(b). In the 3 years' timeframe, when the SME actively targets a reduction in its carbon emission by the end of the first year (strategizing in advance), an overall reduction of ca 20% is possible by the end of the third year. Similarly, in the 5 years' time frame, the timeline for a similar action will be the end of the 2-year timeline that then contributes to ca 22% reduction of $CO_2$ emission by the end of the fourth year. *Such recursive monitoring of strategy could continue to improve the carbon footprint in successive years, highlighting a major achievement of this optimal model. This case provides an example that our model not only complements SDG 13, i.e. climate action, but also addresses SDG 15 which emphasizes the impact of life on land. As an SME makes effort to reduce the $CO_2$ emissions to meet a preset target, it is essentially contributing towards negative climate change inspiring better life sustenance.*

**4.2** ***Case Study 2: The SME chooses to increase the product sale without increasing the logistics cost with a certain compromise on the carbon reduction (i.e. carbon emission decreases)***

This is a unique case where the SME chooses to increase the production and aims to flourish by bringing up the sale and focuses on carbon reduction as well. However, they are not willing to further add up on the logistics cost. Mathematically, this changes the boundary conditions for the associated two variables, together with an increase in the number of operations. In this case, the output variables are associated with SDG 9, 11, 12 and 13.

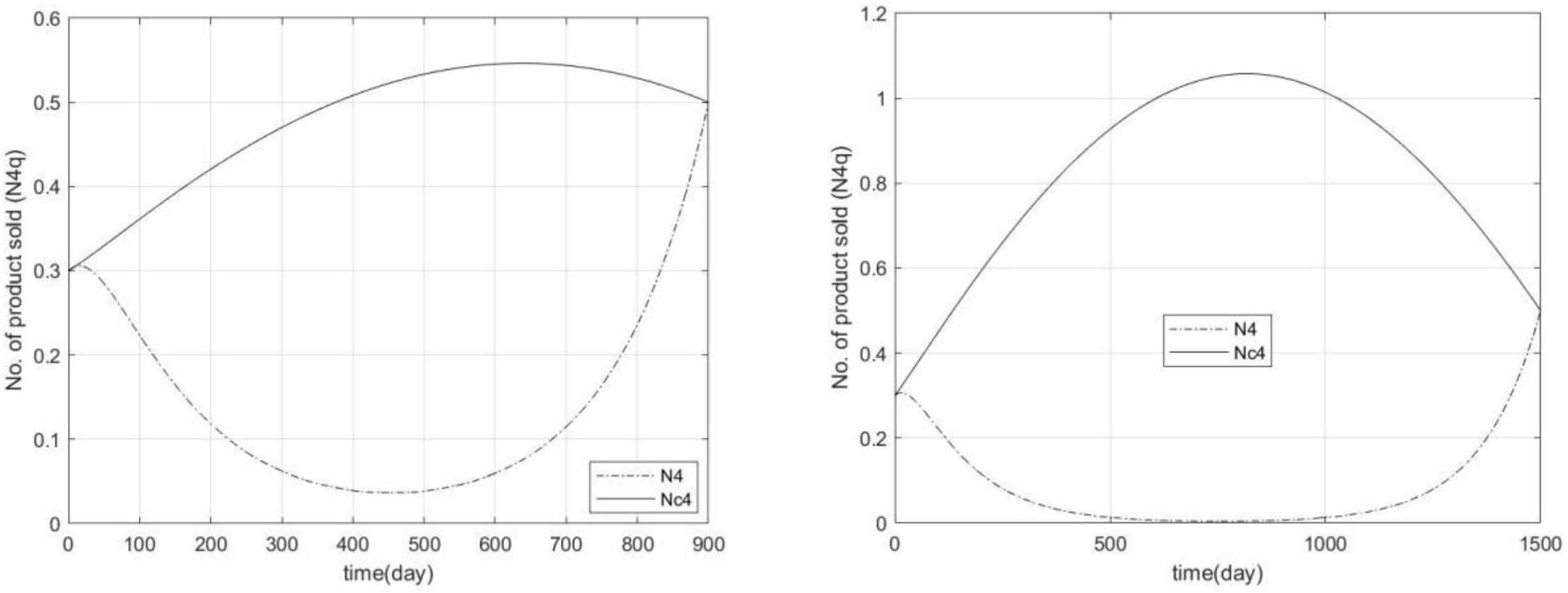

(a) (b)

Figure 7: Time dependency of $N_4$ in unconstrained (dotted line) and constrained (solid line) environment, plots obtained from simultaneous solution of Eqs. (10) and (13): (a) 3-year time span; (b) 5-year time

In the unconstrained environment, the product sale decreases to a minimum of less than 10% during the middle of the second year and eventually recovers to reach the goal in the end of the third year (Figure 7a). Compared to its 3-year strategy, the 5-year strategy shows higher time period of loss which is during $500 < time < 900$ and then exponential recovery to reach goal (Figure 7b). *This suggests that, in arbitrage condition, the strategy might reach the goal, but the SME may not be flourishing, rather it will be operating in a break-even state for a significant amount of time. This demands a change in policy for the SME at a certain interval to reinvigorate product sales.*

In the constrained system, the product sale of the SME increases exponentially and reaches a maximum of 55% at the end of the second year and reaches the goal in the end of the time period of three years. Conversely, in the 5-year strategy, product sale increases and exceeds 100% in the end of the third year. The sale decreases thereafter and reaches the prefixed boundary values. As long as it does not affect the production and economic sustainability of the SME, the strategy may prevail. *Our model clearly identifies the action points for the SME. The maxima of the curves represent time points when the SME must intervene with an appropriate change in* *policy. It is suggested that regular monitoring and cross-validation of existing policy should be carried out for business sustainability.*

It is imperative from the results obtained that increasing product sales will not only increase resource consumption (SDG 12) but will also impact on climate change (SDG 13) due to carbon emissions from the processes involved. This will negatively impact life on land and water (SDG 14 and 15). Our work focuses on the optimization of logistics cost that will part offset this negative input bearing in mind that technologically such emissions are not completely avoidable. As suggested, policy changes and regular monitoring will certainly lead to positive effects towards achieving the SDGs. Sustainable consumption of resources will not only contribute towards positive impacts on SDG 12 – 15; but it will also address certain aspects of sustainable cities (SDG 11). Overall, there are enough scopes for partnership among goals (SDG 17) that can lead to decent economic growth (SDG 8).

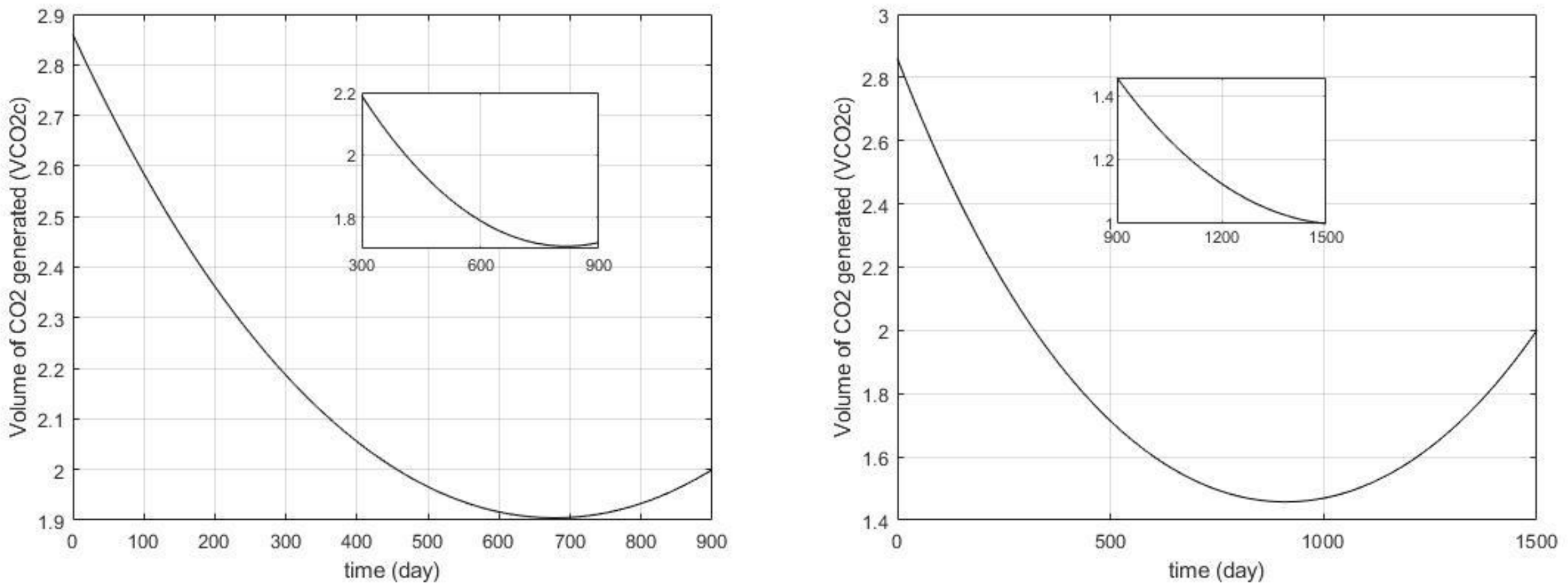

(a) (b)

Figure 8: Time Dependency of $VCO_2$ in a constrained environment, plots obtained from simultaneous solution of Eqs. (10) and (13): (a) 3-year time span & (b) 5-year time span

In this case, the SME takes active initiatives to reduce carbon emission. The emission curve has a parabolic profile and attains a minimum before reaching its desired value both in the 3-year (Figure 8a) and 5-year (Figure 8b) time spans. *This implies that the SME has managed to implement an environmentally sustainable strategy. As mentioned in the preceding subsection of the discussion, there are positive impacts towards climate change (SDG 13) and this strategy will help them to attain the desired goal.*

*Our model suggests that the SME can strategically reduce their carbon footprint by periodically reorienting the strategy after regular time intervals. However, in this case, the model does not specifically constrain a strategic revision at a specific cost (function)-minimum, rather the choice will remain with the SME to periodically monitor and strategize in advance*. The results are provided in the insets of figure 8 (a) and 8 (b). In the 3-year timeframe, when the SME revises its carbon emission at the end of the first year, a reduction of ca 22% is achievable by the end of the third year. Likewise, in the 5-year timeframe, the target timeline for revision will be the end of the third year which makes it possible to reduce carbon emissions to ca 30% by the end of the fifth year. *The alluded case also refers to a situation where both demand uncertainty and environmental uncertainty can be eliminated in a single intervention. Additionally, the general negligence of the SMEs towards carbon footprint reduction can be repaired without compromising the profit margin, as this particular case demonstrates*. Hence, this case offers its flexibility to jointly address SDG 8 and 13 and also SDG 9. The positive impact of this case will lead to sustainable cities (SDG 11) and life on land and under water (SDG 14 and 15).

### *4.3 Discussions from the in line with Cleaner Production and SDGs*

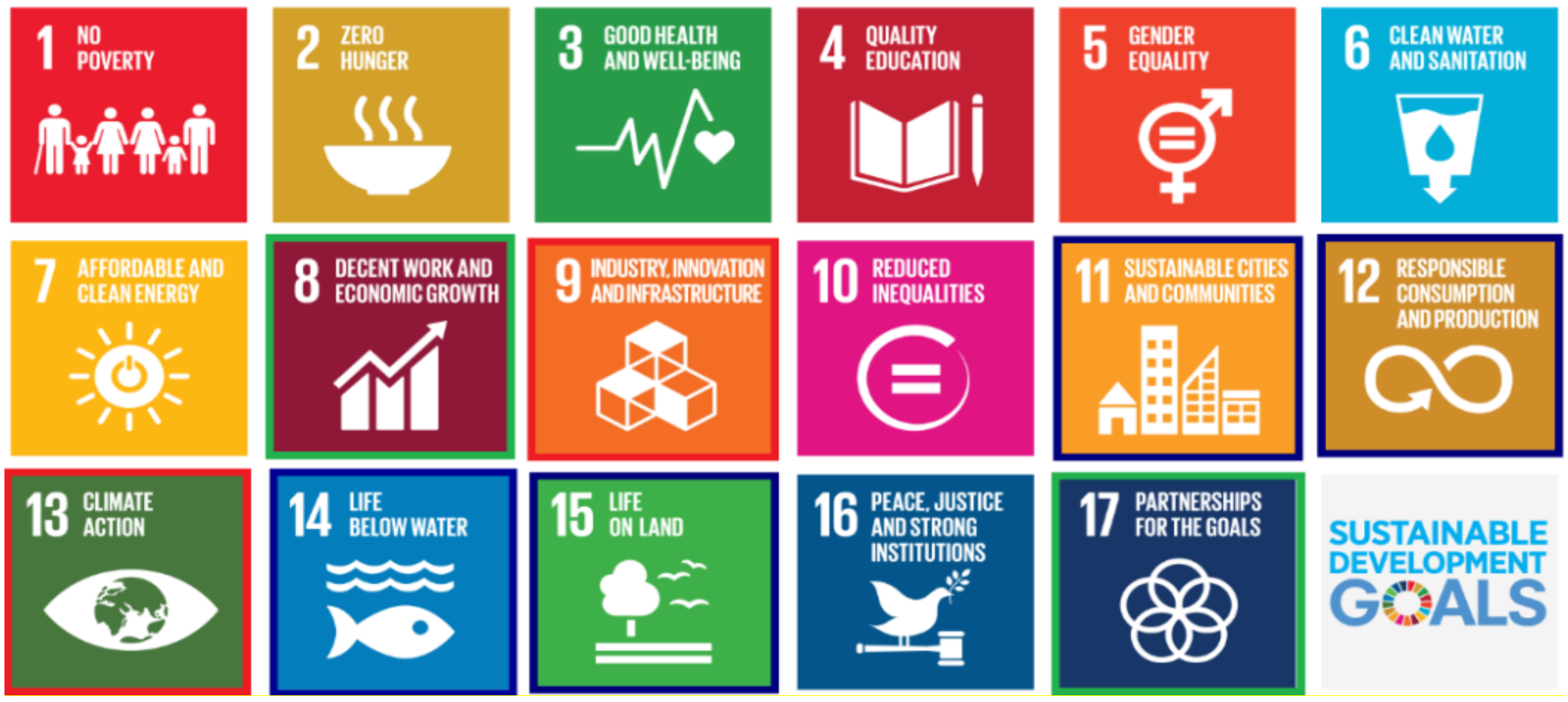

Figure 9: Sustainable Development Goals and their relevance to this work (Source: un.org)

The current investigation, though highly focused on supply chain analytics using uncertainties as metrics, is a generalized structure that any industry can adopt for their sustainability performance measurement and hence move towards a ‘sustainability practice’ approach. The findings of the study largely complement the Sustainable Development Goals (SDGs) 2030 adopted by the UN member countries in 2015. Based on the findings, we have identified eight SDGs which are being addressed by our study. As outlined under Figure 9, our analysis categorizes incumbent factors under three categories - most relevant, moderately relevant and least relevant, respectively identified by color codes red, blue and green.

The two case studies provided outline interesting results where $N_4$ and $VCO_2$ are the key decision making variables. In case study 1, the SME fails to achieve projected target even in favorable conditions. The cleaner production strategy for the SME is to choose their strategies based on market fluctuations and thereby reducing the risk of bankruptcy. From the environmental perspective, the SME achieves nearly 22% reduction in carbon footprint and recursive monitoring is the key to success. CP strategies such as this one will not only enhance the environmental image but also a clean and green societal image for the SME. Case study 2 identifies the action point where CP strategies needs to be implemented. The environmental achievement is substantial and suggests that periodic threshold revisions will provide better results. In summary, our model is able to identify the action points for implementation of CP strategies as well as the SDGs it targets to address. Obviously, the numbers and evolved variables are subjective of the datasets of the current SME but the approach is generic enough for the technique to be ported to other challenging cases such as the waste management sector as well as biorefinery sectors.

**5. Conclusions**

Under the current investigation, a ‘free energy’ model has been developed which considers four groups of uncertainties - Environmental, Economic, Social and Demand. The model is complemented by weight factors and interdependency of the uncertainty factors which makes it robust. The model is parametrized against a range of business weight factors and interdependency terms. Analytic Hierarchical Process (AHP) defines their analytical interdependence. Euler-Lagrange equation is then used to define the time dynamics of the market and its interacting factors. Two cases are developed – a) Unconstrained optimization and b) constrained optimization. Market dissipation and trade friction are then analyzed using Lagrange multipliers that combine the Hamiltonian market force structure, derived from the Euler-Lagrange construct. Further optimization is carried out by solving the boundary value problems using MATLAB and Mathematica.

As an independent test study, we choose data from an anonymous Indian SME to test the robustness of the model. Two cases have been explored, separately in constrained and unconstrained environments, respectively for 3- and 5-year time spans. Results from case 1 show that carbon footprint ignorance may be less consequential for a short period in a constrained environment, but it will eventually drive bankruptcy. The carbon emission increase in this case represents negative growth that is environmentally unsustainable. Advance strategizing toward carbon footprint reduction led to an improved performance. This is a major testament of the strength of this model. Case 2 offers a scenario where sales reinforce carbon reduction targets leading to economic arbitrage where the SME will be operating in a break-even state. The carbon footprint is reduced to a setup target and re-strategizing offers further reduction. This case also suggests that, the time of action is independent of the cost function minimum rather it is a choice that remains with the SME.

The two test cases clearly indicate that right strategy and the correct time of implementation play major role in business prognosis in terms of optimizing cleaner production and profitability priorities. Our model is robust enough to handle diverse streamlined situations, e.g. SME and beyond, a marked improvement on existing practices that are largely limited to objective decision making instead of constrained subjective inputs as our mathematical architecture entails. Future studies involving different data sets from waste management sectors are underway with more generic kernel structure contributed jointly by ANP, Fuzzy and AHP with an aim to deliver a consummate decision-making toolbox.

**Acknowledgement**
The authors acknowledge partial financial supports from the Commonwealth Scholarships Commission (Reference: INCN-2018-52) and Aston University. Dr. Prasanta Kumar Dey, Aston Business School is gratefully acknowledged for his general advice on supply chain literature.

**Contributions**
Led by AKC, the basic framework was developed by AKC, later extended by BD. Numerical computations in this project were conducted by REH, BD and AKC with SME data inputs overseen by SKG; BD conceptualized the AHP results, together with RB; the paper writing was led by AKC, supported by BD, later proof read by SKG.

## Appendix 1

## (Analytic Hierarchy Process - AHP)

### Steps of AHP

The Analytic Hierarchy Process (AHP), developed by T.L. Saaty, was employed to derive the weight factors of the uncertainty variables in the ‘free-energy’ model. The general steps of performing AHP is described below:

*Step 1*: Defining the problem and determining its goal and structuring the hierarchy from the top through the intermediate levels to the lowest level containing the list of alternatives.

*Step 2*: Construction of a set of pair-wise comparison matrices (size n x n) for each of the lower levels with one matrix for each element in the level immediately above by using the relative scale measurement. The pair-wise comparisons are done in terms of which element dominates the other.

*Step 3*: Picked in groups of two over a set of n terms, this produces n(n-1)/2 judgments to develop the set of matrices in step 2. Reciprocals are automatically assigned in each pair-wise comparison.

*Step 4*: Hierarchical synthesis is now used to weight the eigenvectors (normalized representation in the generalized vector space) by the weights of the criteria and the sum is taken over all weighted eigenvector entries corresponding to those in the next lower level of the hierarchy. The AHP eigenvalues represent the normalized weights of the respective quantities, more in the mold of PCA.

*Step 5*: The consistency ratio (CR) confirms the reliability of the pairwise comparisons and it is determined as follows

$$\text{CR} = \frac{\lambda max - n}{n-1} \frac{CI}{RandomConsistency} \ \text{CI/RI} \quad \text{(S1)}$$

Here consistency index CI = ($\lambda_{max}$ – n) / (n –1), where $\lambda_{max}$ is the maximum average value and n is the matrix size. The random consistency index (RI) depends on the value of n. CR is acceptable, if it is ≤ 0.10 to have a consistency level. Beyond this value, the judgment matrix gets inconsistent. To obtain a consistent matrix, judgments should be reviewed and improved.

*Step 6*: Steps 2-5 are performed for all levels in the hierarchy.

**Description of the Layered AHP**

The goal of this AHP is to prioritize the alternatives from the perspective of Uncertainty. The criteria cluster consists of four nodes which are basically the four types of uncertainties considered for the study – Environmental, Economic, Social and Demand. The first layer of the alternatives consists of $VCO_2$, $W_p$, $H_p$ and $W_w$ connected from the Environmental node; $N_1$, $N_2$ and $N_3$ connected from the Social Node. The second layer of alternatives consists of $N_4$, $N_5$, $N_6$ and g connected from the economical node; $N_7$, $N_8$ and M connected from the Demand node. In order to establish the interdependencies among the parameters, the nodes in the second layer has been connected to the associated variables in the first layer. $VCO_2$ is connected to L, $N_5$ and $N_7$; $W_p$ is connected to $N_5$& $N_6$; Hp is connected to L and $N_5$; $W_w$ is connected to L and $N_5$ and $N_3$ is interlinked with $N_4$ and $N_6$. This AHP helps to find out the co-efficient of interdependencies directly from the AHP ($a_1$, $b_2$, $c_3$etc). The compound interdependencies have been taken as product of concerned co-efficient, e.g. value of$a_{23}$ = value of $a_2$ x value of $a_3$. The squared co-efficient has been taken as square root of the concerned co-efficient. For example, the value of $a'_3$ = square root of $a_3$.

**Numerical Values Used for Calculation**

**Table S1: Values Derived using AHP used for calculation**

| Quantity | Value | Quantity | Value |
|---|---|---|---|
| $A_1$ | 0.0797 | $a_1$ | 0.10473 |
| $A_2$ | 0.0178 | $a_2$ | 0.25828 |
| $A_3$ | 0.0488 | $a_3$ | 0.63699 |
| $A_4$ | 0.0586 | $a_{12}$ | 0.02705 |
| $A_5$ | 0.0083 | $a_{23}$ | 0.164522 |
| $A_6$ | 0.0115 | $a_{31}$ | 0.066712 |
| $A_7$ | 0.0115 | $a'_1$ | 0.32362 |
| $A_8$ | 0.0573 | $a'_2$ | 0.508213 |
| $A_9$ | 0.1252 | $a'_3$ | 0.798117 |
| $A_{10}$ | 0.0963 | $b_1$ | 0.75 |
| $A_{11}$ | 0.0386 | $b'_1$ | 0.866025 |
| $A_{12}$ | 0.022 | $c_1$ | 0.5 |
| $A_{13}$ | 0.0773 | $c_2$ | 0.5 |
| $A_{14}$ | 0.1567 | $c_3$ | 0.25 |
| $A_{15}$ | 0.1906 | $c'_1$ | 0.707107 |
| $\varepsilon_1$ | 0.42458 | $c'_2$ | 0.707107 |
| $\varepsilon_2$ | 0.28198 | $d_1$ | 0.5 |
| $\varepsilon_3$ | 0.2132 | $d_2$ | 0.5 |
| $\varepsilon_4$ | 0.08024 | $d_3$ | 0.25 |
| $\lambda_1$ | 0.42458 | $d'_1$ | 0.707107 |
| $\lambda_2$ | 0.28198 | $d'_2$ | 0.707107 |
| $\lambda_3$ | 0.2132 | $\beta_1$ | 0.5 |
| $\lambda_4$ | 0.08024 | $\beta_2$ | 0.5 |
| $\alpha_1$ | 0.25 | $\beta_{12}$ | 0.25 |
| $\alpha_2$ | 0.75 | $\beta'_1$ | 0.707107 |
| $\alpha_{12}$ | 0.1875 | $\beta'_2$ | 0.707107 |
| $\alpha'_1$ | 0.5 | $\gamma$ | 0.1361 |
| $\alpha'_2$ | 0.866025 | | |

**Table S2: Data partially modified from an SME**

| Quantity | Value | Quantity | Value |
|---|---|---|---|
| $f_1$ | 10000 | $N_1$ | 42 |
| $f_2$ | 0 | $N_2$ | 9 |
| $f_3$ | 72000 | $N_3$ | 5 |
| $f_4$ | 60000 | $N_4$ | 15634 |
| $f_5$ | 108000 | $N_5$ | 76 |
| $f_6$ | 200000 | $N_6$ | 8 |
| $f_7$ | 160000 | $N_7$ | 2 |
| $f_8$ | 4500 | $N_8$ | 1 |
| $f_9$ | 0 | y | 0 |
| $f_{10}$ | 180000 | M | 166000 |
| $f_{11}$ | 0 | g | 25000 |
| u (scale factor) | 1/3000 | L | 17000 |

## Appendix 2

## (The Unconstrained Model)

***Free energy model with weight factors***

$$F = \sum_i A_1 V_{CO_2} f_{1_i} + \sum_i A_2 H_{P_i} f_{2_i} + \sum_i A_3 W_{P_i} f_{3_i} + \sum_i A_4 W_{w_i} y_i + \sum_i A_5 L_i + \sum_i A_6 N_{1_i} f_{4_i}$$

$$+ \sum_i A_7 N_{2_i} f_{5_i} + \sum_i A_8 N_{3_i} f_{6_i} + + \sum_i A_9 N_{4_i} f_{7_i} - \sum_i A_{10} N_{5_i} f_{8_i} - \sum_i A_{11} f_{9_i} g_i - \sum_i A_{12} T_i N_{6_i}$$

$$+ \sum_i A_{14} f_{11_i} N_{8_i} + \sum_i A_{15} M_i$$

**Detailed Mathematical Model**

The following is the cost function considering all the parameters and interdependencies.

$$F = \varepsilon_1 \left( \sum_i A_1 V_{CO_2} f_{1_i} + \sum_i A_2 H_{P_i} f_{2_i} + \sum_i A_3 W_{P_i} f_{3_i} + \sum_i A_4 W_{w_i} y_i + \sum A_5 L_i \right) + \varepsilon_2 \left( \sum A_6 N_{1_i} f_{4_i} + \sum_i A_7 N_{2_i} f_{5_i} + \sum_i A_8 N_{3_i} f_{6_i} \right)$$

$$+ \varepsilon_3 \left( \sum_i A_9 N_{4_i} f_{7_i} - \sum_i A_{10} N_{5_i} f_{8_i} - \sum_i A_{11} f_{9_i} g_i - \sum_i A_{12} T_i N_{6_i} \right) + \varepsilon_4 \left( \sum A_{13} f_{10_i} N_{7_i} + \sum_i A_{14} f_{11_i} N_{8_i} + \sum_i A_{15} M_i \right)$$

***Inter-dependency of the uncertainty variables***

The interdependencies has been estimated as two-degree polynomial and represented below

$$V_{CO_2} = a_1L + a_2N_5 + a_3N_7 + a_{12}LN_5 + a_{23}N_5N_7 + a_{31}LN_7 + a_1^{'}L^2 + a_2^{'}N_5^{\;2} + a_3^{'}N_7^{\;2}$$
$$W_P = W_P^O + b_1N_5 + b_2N_5^{\;2}$$
$$H_P = c_1L + c_2N_5 + c_{12}LN_5 + c_1^{'}L^2 + c_2^{'}N_5^{\;2}$$
$$W_w = d_1L + d_2N_5 + d_{12}LN_5 + d_1^{'}L^2 + d_2^{'}N_5^{\;2}$$
$$N_3 = \alpha_1N_4 + \alpha_2N_6 + \alpha_{12}N_4N_6 + \alpha_1^{'}N_4^{\;2} + \alpha_2^{'}N_5^{\;2}$$
$$N_4 = \beta_1N_7 + \beta_2N_8 + \beta_{12}N_7N_8 + \beta_1^{'}N_7^{\;2} + \beta_2^{'}N_8^{\;2}$$
$$N_7 = \gamma V_{CO_2}$$

**First Derivatives**

$$\frac{\partial F}{\partial L} = \left[\varepsilon_1A_1a_1 + \varepsilon_1A_2f_2c_1 + \varepsilon_1A_4yd_1 + \varepsilon_4\gamma A_{13}f_{10}a_1 + \varepsilon_1A_5\right] + \left[\varepsilon_1A_1a_{12} + \varepsilon_4\gamma A_{13}f_{10} + \varepsilon_1A_2f_2c_{12} + \varepsilon_1A_4yd_{12}\right]N_5$$
$$+\left[\varepsilon_1A_1 + \varepsilon_4\gamma A_{13}\right]a_{31}N_7 + 2\left[\varepsilon_1A_1 + \varepsilon_4\gamma A_{13}f_{10} + \varepsilon_1A_2f_2c_1^{'} + \varepsilon_1A_4yd_1^{'}\right]L$$

$$\frac{\partial F}{\partial N_1} = \varepsilon_2A_6f_4 \qquad \frac{\partial F}{\partial N_2} = \varepsilon_2A_7f_5$$

$$\frac{\partial F}{\partial N_3} = \varepsilon_2A_8f_6 + \varepsilon_3\left[\frac{A_9f_7}{\alpha_1 + \alpha_{12}N_6 + 2\alpha_1^{'}N_4} - \frac{A_{12}T}{\alpha_2 + \alpha_{12}N_4 + 2\alpha_2^{'}N_6}\right]$$

$$\frac{\partial F}{\partial N_4} = \varepsilon_3A_9f_7 + \varepsilon_2A_8f_6\left(\alpha_1 + \alpha_{12}N_6 + 2\alpha_1^{'}N_4\right) + \left[\frac{\varepsilon_4A_{13}f_{10}}{\beta_1 + \beta_{12}N_8 + 2\beta_1^{'}N_7} + \frac{\varepsilon_4A_{14}f_{11}}{\beta_2 + \beta_{12}N_7 + 2\beta_2^{'}N_8}\right]$$

$$\frac{\partial F}{\partial N_5} = \left[\varepsilon_1A_1f_1a_2 + \varepsilon_4\gamma A_{13}f_{10}a_2 + \varepsilon_1A_2f_2c_2 + \varepsilon_1A_4yd_2 - \varepsilon_3A_{10}f_8\right] +$$
$$\left[2N_5(\varepsilon_1A_1f_1a_2^{'} + \varepsilon_4\gamma A_{13}f_{10}a_2^{'} + \varepsilon_1A_2f_2c_2^{'} + \varepsilon_1A_4yd_2^{'})\right]$$

$$+\left[\varepsilon_1 A_1 f_1 a_{23} + \varepsilon_4 \gamma A_{13} f_{10} a_{23}\right] N_7 + \left[\varepsilon_1 A_1 f_1 a_{12} + \varepsilon_1 A_2 f_2 c_{12} + \varepsilon_1 A_4 y d_{12} + \varepsilon_4 \gamma A_{13} f_{10} a_{12}\right] L$$

$$\frac{\partial V_{CO_2}}{\partial L} = a_1 + a_{12} N_5 + a_{31} N_7 + 2a_1^{'} L$$

$$\frac{\partial W_P}{\partial N_5} = b_1 + 2b_1^{'} N_5$$

$$\frac{\partial H_P}{\partial L} = c_1 + c_{12} N_5 + 2c_1^{'} L$$

$$\frac{\partial H_P}{\partial N_5} = c_2 + c_{12} L + 2c_2^{'} N_5$$

$$\frac{\partial W_w}{\partial L} = d_1 + d_{12} N_5 + 2d_1^{'} L$$

$$\frac{\partial W_w}{\partial N_5} = d_2 + d_{12} L + 2d_2^{'} N_5$$

$$\frac{\partial F}{\partial N_6} = -\varepsilon_3 A_{12} T + \varepsilon_2 A_8 f_6 \left(\alpha_2 + \alpha_{12} N_4 + 2\alpha_2^{'} N_6\right)$$

$$\frac{\partial F}{\partial N_7} = \varepsilon_1 \left[A_1 f_1 \left(a_3 + a_{23} N_5 + a_{31} L + 2a_3^{'} N_7\right)\right] + \varepsilon_3 \left[A_9 f_7 \left(\beta_1 + \beta_{12} N_8 + 2\beta_1^{'} N_7\right)\right] + \varepsilon_4 A_{13} f_{10}$$

$$\frac{\partial F}{\partial N_8} = \varepsilon_3 A_9 f_7 \left[\left(\beta_2 + \beta_{12} N_7 + 2\beta_2^{'} N_8\right)\right] + \varepsilon_4 A_{14} f_{11}$$

$$\frac{\partial F}{\partial g} = -\varepsilon_3 A_{11} f_9 \qquad \frac{\partial F}{\partial M} = \varepsilon_4 A_{15}$$

**Second derivatives**

$$\frac{\partial^2 F}{\partial V_{CO_2}^2} = \frac{1}{a_2 + a_{23} N_7 + a_{12} L + 2a_2 N_5}\left(\frac{\partial^2 F}{\partial V_{CO_2} \partial N_5}\right) + \frac{1}{a_1 + a_{12} N_5 + a_{31} N_7 + 2a_1 L}\left(\frac{\partial^2 F}{\partial V_{CO_2} \partial N_7}\right)$$

$$\frac{\partial^2 F}{\partial {N_3}^2} = \frac{1}{\alpha_1 + \alpha_{12} N_6 + 2\alpha_1^{'} N_4} \times$$

$$\left( \begin{array}{l} \varepsilon_3 A_9 f_7 + \varepsilon_2 A_8 f_6 \left( \alpha_1 + \alpha_{12} N_6 + 2\alpha_1^{'} N_4 \right) \\ + \left[ \frac{\varepsilon_4 A_{13} f_{10}}{\beta_1 + \beta_{12} N_8 + 2\beta_1^{'} N_7} + \frac{\varepsilon_4 A_{14} f_{11}}{\beta_2 + \beta_{12} N_7 + 2\beta_2^{'} N_8} \right] \end{array} \right)$$

$$+ \frac{1}{\alpha_2 + \alpha_{12} N_4 + 2\alpha_2^{'} N_6} \left( \varepsilon_3 A_{12} T + \varepsilon_2 A_8 f_6 \left( \alpha_2 + \alpha_{12} N_4 + 2\alpha_2^{'} N_6 \right) \right)$$

$$\frac{\partial^2 F}{\partial {N_4}^2} = \left( \alpha_1 + \alpha_{12} N_6 + 2\alpha_1^{'} N_4 \right) \times$$

$$\left( \varepsilon_2 A_8 f_6 + \varepsilon_3 \left[ \frac{A_9 f_7}{\alpha_1 + \alpha_{12} N_6 + 2\alpha_1^{'} N_4} - \frac{A_{12} T}{\alpha_2 + \alpha_{12} N_4 + 2\alpha_2^{'} N_6} \right] \right)$$

$$+ \frac{1}{\beta_1 + \beta_{12} N_8 + 2\beta_1^{'} N_7} \left( \begin{array}{l} \varepsilon_1 \left[ A_1 f_1 \left( a_3 + a_{23} N_5 + a_{31} L + 2a_3^{'} N_7 \right) \right] \\ + \varepsilon_3 \left[ A_9 f_7 \left( \beta_1 + \beta_{12} N_8 + 2\beta_1^{'} N_7 \right) \right] + \varepsilon_4 A_{13} f_{10} \end{array} \right)$$

$$+ \frac{1}{\beta_2 + \beta_{12} N_7 + 2\beta_2^{'} N_8} \left( \varepsilon_3 A_9 f_7 \left[ \left( \beta_2 + \beta_{12} N_7 + 2\beta_2^{'} N_8 \right) \right] + \varepsilon_4 A_{14} f_{11} \right)$$

$$\frac{\partial^2 F}{\partial {N_5}^2} = 2 \left( \varepsilon_1 A_1 f_1 a_2^{'} + \gamma \varepsilon_4 A_{13} f_{10} a_2^{'} + \varepsilon_1 A_2 f_2 c_2^{'} + \varepsilon_1 A_4 y d_2^{'} \right)$$

$$+ \gamma \left( a_2 + a_{23} N_7 + a_{12} L + 2a_2^{'} N_5 \right) \left[ \gamma \varepsilon_4 A_{13} f_{10} a_{23} + \varepsilon_1 A_1 f_1 a_{23} \right]$$

$$\frac{\partial^2 F}{\partial {N_7}^2} = \frac{\varepsilon_1 A_1 f_1 a_{23}}{\gamma \left( a_2 + a_{23} N_7 + a_{12} L + 2a_2^{'} N_5 \right)} + 2 \left( \varepsilon_1 A_1 f_1 a_3^{'} + \varepsilon_3 A_9 f_7 \beta_1^{'} \right)$$

$$\frac{\partial^2 F}{\partial N_3 \partial N_7} = -\frac{2\alpha_1^{'} \varepsilon_3 A_9 f_7 \left( \beta_1 + \beta_{12} N_8 + 2\beta_1^{'} N_7 \right)}{\left( \alpha_1 + \alpha_{12} N_6 + 2\alpha_1^{'} N_4 \right)}$$

$$+ \frac{\alpha_{12} A_{12} T \left( \beta_1 + \beta_{12} N_8 + 2\beta_1^{'} N_7 \right)}{\left( \alpha_2 + \alpha_{12} N_4 + 2\alpha_2^{'} N_6 \right)^2}$$

$$\frac{\partial^2 F}{\partial N_3 \partial V_{CO_2}} = \left( -\frac{\gamma 2\alpha_1^{'} \varepsilon_3 A_9 f_7 \left( \beta_1 + \beta_{12} N_8 + 2\beta_1^{'} N_7 \right)}{\left( \alpha_1 + \alpha_{12} N_6 + 2\alpha_1^{'} N_4 \right)} + \frac{\gamma \alpha_{12} A_{12} T \left( \beta_1 + \beta_{12} N_8 + 2\beta_1^{'} N_7 \right)}{\left( \alpha_2 + \alpha_{12} N_4 + 2\alpha_2^{'} N_6 \right)^2} \right)$$

$$\frac{\partial^2 F}{\partial N_4 \partial N_7} = \left(\gamma\left(a_1 + a_{12}N_5 + a_{31}N_7 + 2a_1^{'}L\right)\right)\times$$

$$\left[\frac{2\beta_1^{'}\varepsilon_4 A_{13} f_{10}}{\left(\beta_1 + \beta_{12}N_8 + 2\beta_1^{'}N_7\right)^2} - \frac{\beta_{12}\varepsilon_4 A_{14} f_{11}}{\left(\beta_2 + \beta_{12}N_7 + 2\beta_2^{'}N_8\right)^2}\right]$$

$$+2\varepsilon_2 A_8 f_6 \alpha_1^{'}\left(\beta_1 + \beta_{12}N_8 + 2\beta_1^{'}N_7\right)$$

$$\frac{\partial^2 F}{\partial N_4 \partial V_{CO_2}} = \frac{1}{\left(a_3 + a_{23}N_5 + a_{31}L + 2a_3^{'}N_7\right)}\cdot\left(\frac{\partial^2 F}{\partial N_4 \partial N_7}\right)$$

$$\frac{\partial^2 F}{\partial N_5 \partial V_{CO_2}} = \left(\frac{1}{\left(a_1 + a_{12}N_5 + a_{31}N_7 + 2a_1^{'}L\right)}\right)\times$$

$$\begin{pmatrix}\left(\varepsilon_1 A_1 a_{12} + \gamma\varepsilon_4 A_{13} f_{10} + \varepsilon_1 A_2 f_2 c_{12} + \varepsilon_1 A_4 y d_{12}\right) \\ +\gamma a_{13}\left(\varepsilon_1 A_1 + \varepsilon_4 A_{13}\gamma\right)\left(a_1 + a_{12}N_5 + a_{31}N_7 + 2a_1^{'}L\right)\end{pmatrix}$$

$$+\left(\frac{1}{\left(a_2 + a_{23}N_7 + a_{12}L + 2a_2^{'}N_5\right)}\right)\times$$

$$\begin{pmatrix}2\left(\varepsilon_1 A_1 f_1 a_2^{'} + \gamma\varepsilon_4 A_{13} f_{10} a_2^{'} + \varepsilon_1 A_2 f_2 c_2^{'} + \varepsilon_1 A_4 y d_2^{'}\right) \\ +\gamma\left(a_2 + a_{23}N_7 + a_{12}L + 2a_2^{'}N_5\right)\left[\gamma\varepsilon_4 A_{13} f_{10} a_{23} + \varepsilon_1 A_1 f_1 a_{23}\right]\end{pmatrix}$$

$$\frac{\partial^2 F}{\partial N_7 \partial V_{CO_2}} = \frac{1}{\left(a_2 + a_{23}N_7 + a_{12}L + 2a_2^{'}N_5\right)}\cdot\left(\frac{\partial^2 F}{\partial N_5 \partial N_7}\right)$$

$$+\frac{1}{\left(a_3 + a_{23}N_5 + a_{31}L + 2a_3^{'}N_7\right)}\cdot\left(\frac{\partial^2 F}{\partial N_5 \partial N_7}\right)$$

$$\frac{\partial^2 F}{\partial N_5 \partial N_7} = \left(\frac{2\varepsilon_1 A_1 f_1 a_2^{'} + \varepsilon_4 \gamma A_{13} f_{10} a_2^{'} + 2\varepsilon_1 A_2 f_2 c_2^{'} + 2A_4 y d_2^{'}}{\gamma\left(a_1 + a_{12}N_5 + a_{31}N_7 + 2a_1^{'}L\right)}\right) + \left[\varepsilon_1 A_1 f_1 a_{23} + \varepsilon_4 A_{13} f_{10} a_{23}\right]$$

$$+\left(\frac{\varepsilon_1 A_1 f_1 a_{12} + \varepsilon_1 A_2 f_2 c_{12} + \varepsilon_1 A_4 y d_{12} + \varepsilon_4 A_{13} f_{10} a_{12}}{\gamma\left(a_1 + a_{12}N_5 + a_{31}N_7 + 2a_1^{'}L\right)}\right)$$

**Second derivatives in terms of the Lagrangian:**

$$\frac{\partial^2 \mathcal{L}}{\partial V_{CO_2}^2} = \frac{1}{a_2 + a_{23}N_7 + a_{12}L + 2a_2N_5} \cdot \left( \frac{\partial^2 \mathcal{L}}{\partial V_{CO_2} \partial N_5} \right) + \frac{1}{a_3 + a_{23}N_5 + a_{13}L + 2a_3N_7} \cdot \left( \frac{\partial^2 \mathcal{L}}{\partial V_{CO_2} \partial N_7} \right)$$

$$\frac{\partial^2 \mathcal{L}}{\partial V_{CO_2} \partial N_5} = \frac{1}{a_1 + a_{12}N_5 + a_{31}N_7 + 2a_1^{'}L} \cdot \begin{bmatrix} \left( \varepsilon_1 A_1 f_1 a_{12} + \gamma \varepsilon_4 A_{13} f_{10} + \varepsilon_1 A_2 f_2 c_{12} + \varepsilon_1 A_4 y d_{12} \right) \\ + \gamma a_{13} \left( \varepsilon_1 A_1 f_1 + \varepsilon_4 A_{13} \gamma f_{10} \right) \cdot \left( a_1 + a_{12}N_5 + a_{31}N_7 + 2a_1^{'}L \right) \end{bmatrix}$$

$$+ \frac{1}{a_2 + a_{32}N_7 + a_{21}LN_7 + 2a_2^{'}N_5} \cdot \begin{bmatrix} 2\left( \varepsilon_1 A_1 f_1 a_2^{'} + \gamma \varepsilon_4 A_{13} f_{10} a_2^{'} + \varepsilon_1 A_2 f_2 c_2^{'} + \varepsilon_1 A_4 y d_2 \right) \\ + \gamma \left( a_2 + a_{32}N_7 + a_{21}LN_7 + 2a_2^{'}N_5 \right) \cdot \left( \gamma \varepsilon_4 A_{13} f_{10} a_{23} + \varepsilon_1 A_1 f_1 a_{23} + 2\varepsilon_1 A_1 f_1 b_2 \right) \end{bmatrix}$$

$$\frac{\partial^2 \mathcal{L}}{\partial V_{CO_2} \partial N_7} = \frac{1}{a_2 + a_{32}N_7 + a_{21}L + 2a_2^{'}N_5} \cdot \left( \frac{\partial^2 \mathcal{L}}{\partial N_5 \partial N_7} \right) + \frac{1}{a_3 + a_{23}N_5 + a_{31}L + 2a_3^{'}N_7} \cdot \left( \frac{\partial^2 \mathcal{L}}{\partial N_7^2} \right)$$

$$\frac{\partial^2 \mathcal{L}}{\partial N_5 \partial N_7} = \left( \frac{2\varepsilon_1 A_1 f_1 a_2^{'} + 2\gamma \varepsilon_4 A_{13} f_{10} a_2^{'} + 2\varepsilon_1 A_2 f_2 c_2^{'} + 2A_4 y d_2}{\gamma \left( a_1 + a_{12}N_5 + a_{31}N_7 + 2a_1^{'}L \right)} \right) + \varepsilon_1 A_1 f_1 a_{13} + \varepsilon_4 A_{13} f_{10} a_{13}$$

$$+ \left( \frac{\varepsilon_1 A_1 f_1 a_{12} + \varepsilon_1 A_2 f_2 c_{12} + \varepsilon_1 A_4 y d_{12} + \gamma \varepsilon_{13} f_{10}}{\gamma \left( a_1 + a_{12}N_5 + a_{31}N_7 + 2a_1^{'}L \right)} \right)$$

$$\frac{\partial^2 \mathcal{L}}{\partial N_7^2} = \left( \frac{\varepsilon_1 A_1 f_1 a_{23}}{\gamma \left( a_2 + a_{32}N_7 + a_{21}LN_7 + 2a_2^{'}N_5 \right)} \right) + 2\varepsilon_1 A_1 f_1 a_3^{'} + 2\varepsilon_3 A_9 f_7 \beta_1^{'}$$

$$\frac{\partial^2 \mathcal{L}}{\partial V_{CO_2} \partial N_7} = \frac{1}{a_2 + a_{23}N_7 + a_{12}L + 2a_2^{'}N_5} \cdot \begin{bmatrix} \left( \frac{2\varepsilon_1 A_1 f_1 a_2^{'} + 2\gamma \varepsilon_4 A_{13} f_{10} a_2^{'} + 2\varepsilon_1 A_2 f_2 c_2^{'} + 2A_4 y d_2}{\gamma \left( a_1 + a_{12}N_5 + a_{31}N_7 + 2a_1^{'}L \right)} \right) + \varepsilon_1 A_1 f_1 a_{13} + \varepsilon_4 A_{13} f_{10} a_{13} \\ + \left( \frac{\varepsilon_1 A_1 f_1 a_{12} + \varepsilon_1 A_2 f_2 c_{12} + \varepsilon_1 A_4 y d_{12} + \gamma \varepsilon_{13} f_{10}}{\gamma \left( a_1 + a_{12}N_5 + a_{31}N_7 + 2a_1^{'}L \right)} \right) \end{bmatrix}$$

$$+ \frac{1}{a_3 + a_{23}N_5 + a_{31}L + 2a_3^{'}N_7} \cdot \left[ \left( \frac{\varepsilon_1 A_1 f_1 a_{23}}{\gamma \left( a_2 + a_{32}N_7 + a_{21}LN_7 + 2a_2^{'}N_5 \right)} \right) + 2\varepsilon_1 A_1 f_1 a_3^{'} + 2\varepsilon_3 A_9 f_7 \beta_1^{'} \right]$$

$$\frac{\partial^2 \mathcal{L}}{\partial V_{CO_2}^2} = \frac{1}{a_2 + a_{23}N_7 + a_{12}L + 2a_2N_5}\left[\begin{array}{l} \frac{1}{a_1 + a_{12}N_5 + a_{31}N_7 + 2a_1^{'}L}\cdot\left\{\begin{array}{l}\left(\varepsilon_1 A_1 f_1 a_{12} + \gamma\varepsilon_4 A_{13} f_{10} + \varepsilon_1 A_2 f_2 c_{12} + \varepsilon_1 A_4 y d_{12}\right) \\ +\gamma a_{13}\left(\varepsilon_1 A_1 f_1 + \varepsilon_4 A_{13}\gamma f_{10}\right)\cdot\left(a_1 + a_{12}N_5 + a_{31}N_7 + 2a_1^{'}L\right)\end{array}\right\} \\ + \frac{1}{a_2 + a_{23}N_7 + a_{12}LN_7 + 2a_2^{'}N_5}\cdot\left\{\begin{array}{l} 2\left(\varepsilon_1 A_1 f_1 a_2^{'} + \gamma\varepsilon_4 A_{13} f_{10} a_2^{'} + \varepsilon_1 A_2 f_2 c_2^{'} + \varepsilon_1 A_4 y d_2\right) \\ +\gamma\left(a_2 + a_{32}N_7 + a_{21}LN_7 + 2a_2^{'}N_5\right)\cdot\left(\gamma\varepsilon_4 A_{13} f_{10} a_{23} + \varepsilon_1 A_1 f_1 a_{23} + 2\varepsilon_1 A_1 f_1 b_2\right)\end{array}\right\}\end{array}\right]$$

$$+\frac{1}{a_3 + a_{23}N_5 + a_{13}L + 2a_3N_7}\cdot\left[\begin{array}{l} \frac{1}{a_2 + a_{23}N_7 + a_{12}L + 2a_2^{'}N_5}\cdot\left\{\begin{array}{l}\left(\frac{2\varepsilon_1 A_1 f_1 a_2^{'} + 2\gamma\varepsilon_4 A_{13} f_{10} a_2^{'} + 2\varepsilon_1 A_2 f_2 c_2^{'} + 2A_4 y d_2^{'}}{\gamma\left(a_1 + a_{12}N_5 + a_{31}N_7 + 2a_1^{'}L\right)}\right) + \varepsilon_1 A_1 f_1 a_{13} + \varepsilon_4 A_{13} f_{10} a_{13} \\ + \left(\frac{\varepsilon_1 A_1 f_1 a_{12} + \varepsilon_1 A_2 f_2 c_{12} + \varepsilon_1 A_4 y d_{12} + \gamma\varepsilon_{13} f_{10}}{\gamma\left(a_1 + a_{12}N_5 + a_{31}N_7 + 2a_1^{'}L\right)}\right)\end{array}\right\} \\ + \frac{1}{a_3 + a_{23}N_5 + a_{31}L + 2a_3^{'}N_7}\cdot\left\{\left(\frac{\varepsilon_1 A_1 f_1 a_{23}}{\gamma\left(a_2 + a_{23}N_7 + a_{12}L + 2a_2^{'}N_5\right)}\right) + 2\varepsilon_1 A_1 f_1 a_3^{'} + 2\varepsilon_3 A_9 f_7 \beta_1^{'}\right\}\end{array}\right]$$

$$\frac{\partial^2 \mathcal{L}}{\partial N_5^2} = 2\left(\varepsilon_1 A_1 f_1 a_2^{'} + \gamma\varepsilon_4 A_{13} f_{10} a_2^{'} + \varepsilon_1 A_2 f_2 c_2^{'} + A_4 y d_2^{'}\right) + \gamma\left(a_2 + a_{23}N_7 + a_{12}L + 2a_2^{'}N_5\right).\left(\gamma\varepsilon_4 A_{13} f_{10} a_{23} + \varepsilon_1 A_1 f_1 a_{23}\right) + 2\varepsilon_1 A_3 f_3 b_2$$

$$\frac{\partial^2 \mathcal{L}}{\partial N_4^2} = \left(\alpha_1 + \alpha_{12} N_6 + 2\alpha_1^{'} N_4\right) \cdot \left\{ \varepsilon_2 A_8 f_6 + \varepsilon_3 \cdot \left( \frac{A_9 f_7}{\alpha_1 + \alpha_{12} N_6 + 2\alpha_1^{'} N_4} - \frac{A_{12} T}{\alpha_2 + \alpha_{12} N_4 + 2\alpha_2^{'} N_6} \right) \right\}$$

$$+ \frac{1}{\beta_1 + \beta_{12} N_8 + 2\beta_1^{'} N_7} \cdot \left[ \varepsilon_1 \left\{ A_1 f_1 \left( a_3 + a_{23} N_5 + a_{31} L + 2a_3^{'} N_7 \right) \right\} + \varepsilon_3 \left\{ A_9 f_7 \left( \beta_1 + \beta_{12} N_8 + 2\beta_1^{'} N_7 \right) \right\} + \varepsilon_4 A_{13} f_{10} \right]$$

$$+ \frac{1}{\beta_2 + \beta_{12} N_7 + 2\beta_2^{'} N_8} \cdot \left[ \varepsilon_3 \left\{ A_9 f_7 \left( \beta_1 + \beta_{12} N_8 + 2\beta_1^{'} N_7 \right) \right\} + \varepsilon_4 A_{14} f_{11} \right]$$

$$\frac{\partial^2 \mathcal{L}}{\partial N_4 \partial V_{CO_2}} = \frac{1}{a_2 + a_{23} N_7 + a_{12} L + 2a_2^{'} N_5} \cdot \left( \frac{\partial \mathcal{L}^2}{\partial N_4 \partial N_7} \right)$$

$$\frac{\partial^2 \mathcal{L}}{\partial N_4 \partial N_7} = \gamma \left( a_1 + a_{12} N_5 + a_{31} N_7 + 2a_1^{'} L \right) \cdot \left( \frac{2\beta_1^{'} \varepsilon_4 A_{13} f_{10}}{\beta_1 + \beta_{12} N_8 + 2\beta_1^{'} N_7} - \frac{\beta_{12} \varepsilon_4 A_{14} f_{11}}{\beta_2 + \beta_{12} N_7 + 2\beta_2^{'} N_8} \right)$$

$$+ 2\varepsilon_2 A_8 f_6 \alpha_1^{'} \cdot \left( \beta_1 + \beta_{12} N_8 + 2\beta_1^{'} N_7 \right)$$

$$\frac{\partial^2 \mathcal{L}}{\partial N_4 \partial V_{CO_2}} = \frac{1}{a_2 + a_{23} N_7 + a_{12} L + 2a_2^{'} N_5} \cdot \left[ \begin{array}{l} \gamma \left( a_1 + a_{12} N_5 + a_{31} N_7 + 2a_1^{'} L \right) \cdot \left( \frac{2\beta_1^{'} \varepsilon_4 A_{13} f_{10}}{\beta_1 + \beta_{12} N_8 + 2\beta_1^{'} N_7} - \frac{\beta_{12} \varepsilon_4 A_{14} f_{11}}{\beta_2 + \beta_{12} N_7 + 2\beta_2^{'} N_8} \right) \\ + 2\varepsilon_2 A_8 f_6 \alpha_1^{'} \cdot \left( \beta_1 + \beta_{12} N_8 + 2\beta_1^{'} N_7 \right) \end{array} \right]$$

$$
H = \begin{bmatrix}
\frac{\partial^2 F}{\partial V_{CO_2}^2} & \frac{\partial^2 F}{\partial V_{CO_2}\partial N_1} & \frac{\partial^2 F}{\partial V_{CO_2}\partial N_2} & \frac{\partial^2 F}{\partial V_{CO_2}\partial N_3} & \frac{\partial^2 F}{\partial V_{CO_2}\partial N_4} & \frac{\partial^2 F}{\partial V_{CO_2}\partial N_5} & \frac{\partial^2 F}{\partial V_{CO_2}\partial N_6} & \frac{\partial^2 F}{\partial V_{CO_2}\partial N_7} & \frac{\partial^2 F}{\partial V_{CO_2}\partial N_8} & \frac{\partial^2 F}{\partial V_{CO_2}\partial H_P} & \frac{\partial^2 F}{\partial V_{CO_2}\partial W_w} \\
\frac{\partial^2 F}{\partial N_1\partial V_{CO_2}} & \frac{\partial^2 F}{\partial N_1^2} & \frac{\partial^2 F}{\partial N_1\partial N_2} & \frac{\partial^2 F}{\partial N_1\partial N_3} & \frac{\partial^2 F}{\partial N_1\partial N_4} & \frac{\partial^2 F}{\partial N_1\partial N_5} & \frac{\partial^2 F}{\partial N_1\partial N_6} & \frac{\partial^2 F}{\partial N_1\partial N_7} & \frac{\partial^2 F}{\partial N_1\partial N_8} & \frac{\partial^2 F}{\partial N_1\partial W_P} & \frac{\partial^2 F}{\partial N_1\partial H_P} & \frac{\partial^2 F}{\partial N_1\partial W_w} \\
\frac{\partial^2 F}{\partial N_2\partial V_{CO_2}} & \frac{\partial^2 F}{\partial N_2\partial N_1} & \frac{\partial^2 F}{\partial N_2^2} & \frac{\partial^2 F}{\partial N_2\partial N_3} & \frac{\partial^2 F}{\partial N_2\partial N_4} & \frac{\partial^2 F}{\partial N_2\partial N_5} & \frac{\partial^2 F}{\partial N_2\partial N_6} & \frac{\partial^2 F}{\partial N_2\partial N_7} & \frac{\partial^2 F}{\partial N_2\partial N_8} & \frac{\partial^2 F}{\partial N_2\partial W_P} & \frac{\partial^2 F}{\partial N_2\partial H_P} & \frac{\partial^2 F}{\partial N_2\partial W_w} \\
\frac{\partial^2 F}{\partial N_3\partial V_{CO_2}} & \frac{\partial^2 F}{\partial N_3\partial N_1} & \frac{\partial^2 F}{\partial N_3\partial N_2} & \frac{\partial^2 F}{\partial N_3^2} & \frac{\partial^2 F}{\partial N_3\partial N_4} & \frac{\partial^2 F}{\partial N_3\partial N_5} & \frac{\partial^2 F}{\partial N_3\partial N_6} & \frac{\partial^2 F}{\partial N_3\partial N_7} & \frac{\partial^2 F}{\partial N_3\partial N_8} & \frac{\partial^2 F}{\partial N_3\partial W_P} & \frac{\partial^2 F}{\partial N_3\partial H_P} & \frac{\partial^2 F}{\partial N_3\partial W_w} \\
\frac{\partial^2 F}{\partial N_4\partial V_{CO_2}} & \frac{\partial^2 F}{\partial N_4\partial N_1} & \frac{\partial^2 F}{\partial N_4\partial N_2} & \frac{\partial^2 F}{\partial N_4\partial N_3} & \frac{\partial^2 F}{\partial N_4^2} & \frac{\partial^2 F}{\partial N_4\partial N_5} & \frac{\partial^2 F}{\partial N_4\partial N_6} & \frac{\partial^2 F}{\partial N_4\partial N_7} & \frac{\partial^2 F}{\partial N_4\partial N_8} & \frac{\partial^2 F}{\partial N_4\partial W_P} & \frac{\partial^2 F}{\partial N_4\partial H_P} & \frac{\partial^2 F}{\partial N_4\partial W_w} \\
\frac{\partial^2 F}{\partial N_5\partial V_{CO_2}} & \frac{\partial^2 F}{\partial N_5\partial N_1} & \frac{\partial^2 F}{\partial N_5\partial N_2} & \frac{\partial^2 F}{\partial N_5\partial N_3} & \frac{\partial^2 F}{\partial N_5\partial N_4} & \frac{\partial^2 F}{\partial N_5^2} & \frac{\partial^2 F}{\partial N_5\partial N_6} & \frac{\partial^2 F}{\partial N_5\partial N_7} & \frac{\partial^2 F}{\partial N_5\partial N_8} & \frac{\partial^2 F}{\partial N_5\partial W_P} & \frac{\partial^2 F}{\partial N_5\partial H_P} & \frac{\partial^2 F}{\partial N_5\partial W_w} \\
\frac{\partial^2 F}{\partial N_6\partial V_{CO_2}} & \frac{\partial^2 F}{\partial N_6\partial N_1} & \frac{\partial^2 F}{\partial N_6\partial N_2} & \frac{\partial^2 F}{\partial N_6\partial N_3} & \frac{\partial^2 F}{\partial N_6\partial N_4} & \frac{\partial^2 F}{\partial N_6\partial N_5} & \frac{\partial^2 F}{\partial N_6^2} & \frac{\partial^2 F}{\partial N_6\partial N_7} & \frac{\partial^2 F}{\partial N_6\partial N_8} & \frac{\partial^2 F}{\partial N_6\partial W_P} & \frac{\partial^2 F}{\partial N_6\partial H_P} & \frac{\partial^2 F}{\partial N_6\partial W_w} \\
\frac{\partial^2 F}{\partial N_7\partial V_{CO_2}} & \frac{\partial^2 F}{\partial N_7\partial N_1} & \frac{\partial^2 F}{\partial N_7\partial N_2} & \frac{\partial^2 F}{\partial N_7\partial N_3} & \frac{\partial^2 F}{\partial N_7\partial N_4} & \frac{\partial^2 F}{\partial N_7\partial N_5} & \frac{\partial^2 F}{\partial N_7\partial N_6} & \frac{\partial^2 F}{\partial N_7^2} & \frac{\partial^2 F}{\partial N_7\partial N_8} & \frac{\partial^2 F}{\partial N_7\partial W_P} & \frac{\partial^2 F}{\partial N_7\partial H_P} & \frac{\partial^2 F}{\partial N_7\partial W_w} \\
\frac{\partial^2 F}{\partial N_8\partial V_{CO_2}} & \frac{\partial^2 F}{\partial N_8\partial N_1} & \frac{\partial^2 F}{\partial N_8\partial N_2} & \frac{\partial^2 F}{\partial N_8\partial N_3} & \frac{\partial^2 F}{\partial N_8\partial N_4} & \frac{\partial^2 F}{\partial N_8\partial N_5} & \frac{\partial^2 F}{\partial N_8\partial N_6} & \frac{\partial^2 F}{\partial N_8\partial N_7} & \frac{\partial^2 F}{\partial N_8^2} & \frac{\partial^2 F}{\partial N_8\partial W_P} & \frac{\partial^2 F}{\partial N_8\partial H_P} & \frac{\partial^2 F}{\partial N_8\partial W_w} \\
\frac{\partial^2 F}{\partial W_P\partial V_{CO_2}} & \frac{\partial^2 F}{\partial W_P\partial N_1} & \frac{\partial^2 F}{\partial W_P\partial N_2} & \frac{\partial^2 F}{\partial W_P\partial N_3} & \frac{\partial^2 F}{\partial W_P\partial N_4} & \frac{\partial^2 F}{\partial W_P\partial N_5} & \frac{\partial^2 F}{\partial W_P\partial N_6} & \frac{\partial^2 F}{\partial W_P\partial N_7} & \frac{\partial^2 F}{\partial W_P\partial N_8} & \frac{\partial^2 F}{\partial W_P^2} & \frac{\partial^2 F}{\partial W_P\partial H_P} & \frac{\partial^2 F}{\partial W_P\partial W_w} \\
\frac{\partial^2 F}{\partial H_P\partial V_{CO_2}} & \frac{\partial^2 F}{\partial H_P\partial N_1} & \frac{\partial^2 F}{\partial H_P\partial N_2} & \frac{\partial^2 F}{\partial H_P\partial N_3} & \frac{\partial^2 F}{\partial H_P\partial N_4} & \frac{\partial^2 F}{\partial H_P\partial N_5} & \frac{\partial^2 F}{\partial H_P\partial N_6} & \frac{\partial^2 F}{\partial H_P\partial N_7} & \frac{\partial^2 F}{\partial H_P\partial N_8} & \frac{\partial^2 F}{\partial H_P\partial W_P} & \frac{\partial^2 F}{\partial H_P^2} & \frac{\partial^2 F}{\partial H_P\partial W_w} \\
\frac{\partial^2 F}{\partial W_w\partial V_{CO_2}} & \frac{\partial^2 F}{\partial W_w\partial N_1} & \frac{\partial^2 F}{\partial W_w\partial N_2} & \frac{\partial^2 F}{\partial W_w\partial N_3} & \frac{\partial^2 F}{\partial W_w\partial N_4} & \frac{\partial^2 F}{\partial W_w\partial N_5} & \frac{\partial^2 F}{\partial W_w\partial N_6} & \frac{\partial^2 F}{\partial W_w\partial N_7} & \frac{\partial^2 F}{\partial W_w\partial N_8} & \frac{\partial^2 F}{\partial W_w\partial W_P} & \frac{\partial^2 F}{\partial W_w\partial H_P} & \frac{\partial^2 F}{\partial W_w^2}
\end{bmatrix}
$$

The above H-matrix leads to the constrained 4x4 Hessian as below (details in the main text):

$$
\mathrm{H} = \begin{bmatrix}
\frac{\partial^2 \mathcal{L}}{\partial V_{CO_2}^2} & 0 & 0 & 0 \\
\frac{\partial^2 \mathcal{L}}{\partial N_4 \partial V_{CO_2}} & \frac{\partial^2 \mathcal{L}}{\partial N_4^2} & 0 & \frac{\partial^2 \mathcal{L}}{\partial N_4 \partial N_7} \\
\frac{\partial^2 \mathcal{L}}{\partial N_5 \partial V_{CO_2}} & 0 & \frac{\partial^2 \mathcal{L}}{\partial N_5^2} & \frac{\partial^2 \mathcal{L}}{\partial N_5 \partial N_7} \\
\frac{\partial^2 \mathcal{L}}{\partial N_7 \partial V_{CO_2}} & 0 & 0 & \frac{\partial^2 \mathcal{L}}{\partial N_7^2}
\end{bmatrix}
$$

**Elements of Matrix m**

$M_{11} = K_{11} = 2\epsilon_2 A_8 f_6 \alpha_1'$

$$M_{13} = K_{13} = -\epsilon_4 A_{13} f_{10} \frac{2\beta_1'}{(\beta_1 + \beta_{12} N_8^0 + 2\beta_1' N_7^0)^2} - \epsilon_4 A_{14} f_{11} \frac{\beta_{12}}{(\beta_1 + \beta_{12} N_7^0 + 2\beta_2' N_8^0)^2}$$

$M_{22} = K_{22} = 2(\epsilon_1 A_1 f_1 a_2' + \gamma\epsilon_4 A_{13} f_{10} a_2' + \epsilon_1 A_2 f_2 c_2' + \epsilon_1 A_4 y d_2' + \epsilon_1 A_3 f_3 b_2)$

$M_{23} = K_{23} = (\epsilon_1 A_1 f_1 a_{23} + \gamma\epsilon_4 A_{13} f_{10} a_{23})$

$M_{32} = K_{32} = \epsilon_1 A_1 f_1 a_{23}$

$M_{33} = K_{33} = 2\epsilon_1 A_1 f_1 a_3' + 2\epsilon_3 A_9 f_7 \beta_1'$

$$M_{42} = K_{42} = \frac{-a_{12}}{(a_1 + a_{12} N_5 + a_{31} N_7 + 2a_1' L)^2} [(\epsilon_1 A_1 f_1 a_{12} + \gamma\epsilon_4 A_{13} f_{10} + \epsilon_1 A_2 f_2 c_{12} + \epsilon_1 A_4 y d_{12}) N_5 + (\epsilon_1 A_1 f_1 a_{31} + \gamma\epsilon_4 A_{13} f_{10} a_{31}) N_7 + 2(\epsilon_1 A_1 f_1 a_1' + \gamma\epsilon_4 A_{13} f_{10} a_1' + \epsilon_1 A_2 f_2 c_1' + \epsilon_1 A_4 y d_1') L] + \frac{1}{(a_1 + a_{12} N_5 + a_{31} N_7 + 2a_1' L)}$$

$M_{44} = K_{44} = -\lambda_3 f_1$

## Appendix 3

## (Results of Unconstrained Optimization for $VCO_2$)

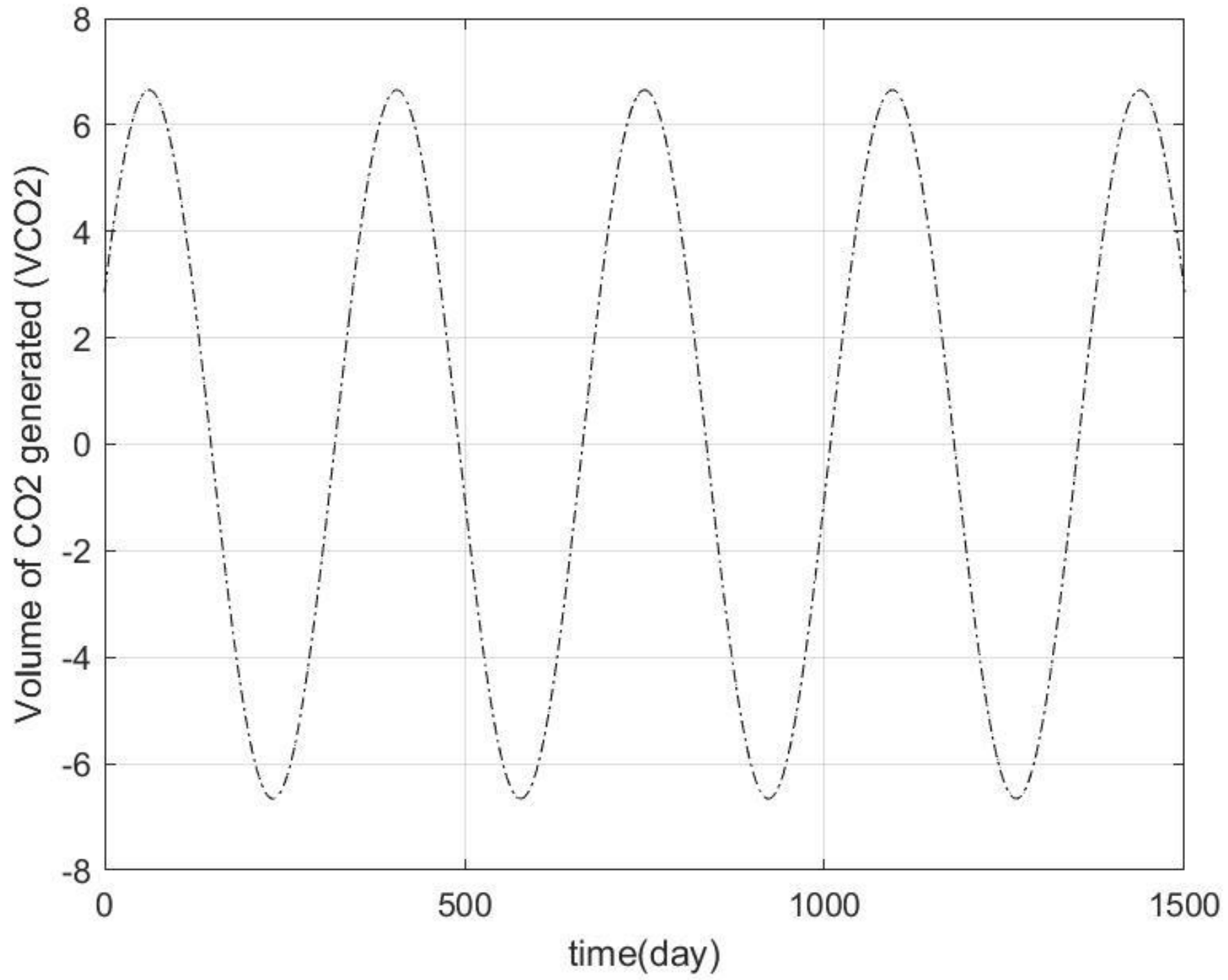

Figure S1: Time Dependency of $VCO_2$ in a constrained environment: 5-year time span

As mentioned in section 5.1, $V_{CO2}$ gives non-converging solutions. The time dependent plot of $VCO_2$ in an unconstrained environment is presented in Figure S1. The curve has an oscillatory profile which implies that in the unconstrained environment, the system is unstable without any control of carbon emission.

## SUPPLEMENTARY READING

Table S3: Variables contributing to uncertainty along the supply chain network

| Sl. | Symbols | Uncertainty Variable | Sources |
|---|---|---|---|
| 1 | $V_{CO2}$ | Volume of $CO_2$ generated | Ahn and Han 2018 |
| 2 | $H_p$ | Thermal Pollution due to processes involved | Firoz et al. 2018 |
| 3 | $W_p$ | Water used due to the processes involved | Guerra et al. 2019 |
| 4 | $W_w$ | Waste water produced in the whole process | Ekşioğlu et al. 2013 |
| 5 | $N_1$ | No. of labours | Bhatnagar and Sohal 2005 |
| 6 | $N_2$ | No. of employees | Fuentes-Fuentes et al., 2004 |
| 7 | $N_3$ | No. of CSR activities in a year | Walker and Jones 2012 |
| 8 | $N_4$ | No. of products sold | Chen and Lee 2004 |
| 9 | $N_5$ | No. of operations involved | Santoso et al. 2005 |
| 11 | $N_6$ | No. of type of taxes | Huh and Park 2013 |
| 12 | $N_7$ | No. of transportations involved | Sanchez Rodrigues et al. 2008 |
| 13 | $N_8$ | No. of other logistics | Synthetic data |
| 14 | $f_1$ | Unit cost for $CO_2$ recovery | Fleten et al. 2010 |
| 15 | $f_2$ | Unit cost for thermal pollution prevention | This study |
| 16 | $f_3$ | Unit cost for water used | Gao and You 2015 |
| 17 | $f_4$ | Wage of one labour | Bhatnagar and Sohal 2005 |
| 18 | $f_5$ | Salary of one employee | Fuentes-Fuentes et al., 2004 |
| 19 | $f_6$ | Average unit cost CSR activity | Carter & Jennings 2002 |
| 20 | $f_7$ | Unit revenue earned from product sold | Heydari & Ghasemi 2018 |
| 21 | $f_8$ | Unit cost of each operation | Guillén et al. 2005 |
| 22 | $f_9$ | Average cost for disaster management per day | Ergan et al. 2010 |
| 23 | $f_{10}$ | Unit cost for transportation | Haddadsisakht & Ryan 2018 |
| 24 | $f_{11}$ | Unit cost for other logistics | Kim et al. 2018 |
| 25 | y | Unit cost for waste water treatment | Ekşioğlu et al. 2013 |
| 26 | M | Miscellaneous cost | This Study |
| 27 | L | Legislative Costs | Rahman & Subramanian 2012 |
| 28 | T | Unit cost for Taxes | Guillén et al. 2005 |
| 29 | g | Disaster management fund | Ergan et al. 2010 |
| 30 | i | Iteration counter | Pishvaee et al. 2011 |